\documentclass[fleqn,usenatbib]{mnras}
\usepackage{newtxtext,newtxmath}
\usepackage[T1]{fontenc}
\usepackage{multicol}
\usepackage{multirow}
\usepackage{natbib}
\usepackage{graphicx}
\usepackage{amsmath}
\usepackage{xcolor}
\usepackage{upgreek}

\title[Evolved hypergiants in the Magellanic Clouds]{Revisiting the evolved hypergiants in the Magellanic Clouds}

\author[Kourniotis et al.]{
M. Kourniotis$^{1}$\thanks{E-mail: kourniotis@asu.cas.cz},
M. Kraus$^{2}$, 
O. Maryeva$^{2}$, 
M. Borges Fernandes$^{3}$,
and G. Maravelias$^{4, 5}$\\
$^{1}$Astronomical Institute, Czech Academy of Sciences, Bocni II 1401, 141 31 Prague, Czech Republic\\
$^{2}$Astronomical Institute, Czech Academy of Sciences, 251\,65 Ond\v{r}ejov, Czech Republic\\
$^{3}$Observat\'orio Nacional, Rua General Jos\'e Cristino 77, 20921-400,
Rio de Janeiro, Brazil\\
$^{4}$Institute for Astronomy, Astrophysics, Space Applications and Remote Sensing, National Observatory of Athens, GR-15236, Penteli, Athens, Greece\\
$^{5}$Institute of Astrophysics, Foundation for Research and Technology-Hellas, GR 71110 Heraklion, Greece} 

\date{Accepted 2022 February 8. Received 2022 January 31; in original form 2021 December 16}
\pubyear{2022}

\begin{document}
\label{firstpage}
\pagerange{\pageref{firstpage}--\pageref{lastpage}}
\maketitle

\begin{abstract}

The massive stars that survive the phase of red supergiants (RSGs) spend the rest of their life in extremity. Their unstable atmospheres facilitate the formation and episodic ejection of shells that alter the stellar appearance and surroundings. In the present study, we revise the evolutionary state of eight hypergiants in the Magellanic Clouds, four of early-A type and four of FG type, and complement the short list of the eruptive post-RSGs termed as yellow hypergiants (YHGs). We refine the outdated temperatures and luminosities of the stars by means of high-resolution spectroscopy with FEROS. The A-type stars are suggested to be in their early, post-main sequence phase, showing spectrophotometric characteristics of redward evolving supergiants. On the other hand, the FG-type stars manifest themselves through the enhanced atmospheric activity that is traced by emission filling in H$\alpha$ and the dynamical modulation of the low-excitation \ion{Ba}{II} line. Of these stars, the dusty HD269723 is suggested to have recently departed from a cool phase. We identify double-peaked emission in the FEROS data of HD269953 that emerges from an orbiting disk-hosting companion. The highlight of the study is an episode of enhanced mass loss of HD271182 that manifests as a dimming event in the lightcurve and renders the star ``modest’’ analogue to $\rho$ Cas. The luminosity $\log(L/L_{\odot})= 5.6$ of HD271182 can serve as an updated threshold for the luminosity of stars exhibiting a post-RSG evolution in the Large Magellanic Cloud.

\end{abstract}


\section{Introduction}

The upper limit of 17$-$20 M$_\odot$ for the initial mass of the type II-P supernova (SN) progenitors \citep{2009ARA&A..47...63S} has triggered an ongoing discussion with respect to the puzzling existence of red supergiants (RSGs) with a higher mass \citep[e.g.][]{2006ApJ...645.1102L,2012ApJ...750...97D}. This discrepancy, termed as the ``RSG problem'', is resolved by scenaria such as the underrated extinction of the SN progenitors \citep{2012MNRAS.419.2054W} and the direct collapse into black holes of the most massive RSGs \citep[so called ``failed SNe'';][]{2008ApJ...684.1336K}. The prevalent scenario based on which the theoretical paths are carved \citep[e.g.][]{2012A&A...537A.146E} is that 
the large mass fraction of the helium core during the RSG phase along with the pulsation-driven high mass-loss rates \citep{2010ApJ...717L..62Y} allows stars with M$_{\text{ini}}=20-40$ M$_\odot$ to shed their hydrogen envelope and evolve towards hotter temperatures \citep{2011BSRSL..80..266M}. Consequently, the most luminous RSGs will end their life \textit{presumably} in the blue region of the evolutionary diagram as Wolf-Rayet stars or Luminous Blue Variables (LBVs) \citep{2019Galax...7...92G}. The study by \cite{Groh14}, on the other hand, which pointed out a yellow post-RSG as a candidate progenitor of SN 2013cu, leaves open the possibility of the terminal state of these stars being cooler.

Yellow hypergiants (YHGs) are synonymous to the luminous post-RSGs with a spectral type F$-$G that have already experienced significant loss of their initial mass \citep{deJag98}. The atmospheres of YHGs are very extended showing large-scale motions that assign the spectral lines a strong macroturbulent broadening. An excess line broadening can be also ascribed to non-radial pulsations \citep{1994A&A...291..226L}, the properties of which are variable and weakly constrained \citep{2019A&A...631A..48V}. The hallmark of the class is the remarkable variability in the spectral type that follows large eruption events \citep[e.g.][]{2006AJ....131.2105H}. These are severe mass-loss episodes that trigger the formation of cool pseudo-photospheres; dense and optically thick gas shells, which can veil the actual photosphere of the star \citep{2000A&A...353..163N, 2013A&A...551A..69O}. Such events are mainly found to take place when the YHG displays temperatures of $6\,000-7\,000$ K, the border of an area on the evolutionary diagram that, in the literature of post-RSG evolution, is labelled as the ``Yellow Void'' \citep[YV;][]{1995A&A...302..811N,deJager97}.The occurrence of the high mass-loss episodes is currently believed to rely on dynamical instability in the atmosphere that is paired with the release of recombination energy from partially ionized hydrogen \citep[e.g.][]{lob03}. On the other hand, the assumption of adiabatic instability is reported to be erroneous for the
envelopes of LBVs \citep{1998MNRAS.295..251G} and even, of YHGs (Glatzel et al. in prep.). To date, the mechanism that powers the ejection of mass is still kept in the shadow and may relate to the non-linear growth and evolution of non-adiabatic processes.

The number of YHGs reported in the literature is less than ten, with the well studied ones comprising almost half. Being subject of monitoring for years to several decades, these objects have been investigated for their instability and nature of pulsations \citep[e.g.][]{2000A&A...353..163N,2019A&A...631A..48V}, the geometry and composition of the ejected material \citep{Davies07,2013A&A...551A..69O,2020A&A...635A.183K}, and the presence of companion stars \citep{2014A&A...563A..71C}. The distinctive spectroscopic profile of YHGs has served as template for suggesting new  analogues of the class \citep[e.g.][]{2014A&A...561A..15C}. To date, one of the brightest stars observed in the Galaxy, the iconic $\rho$ Cas, constitutes the cornerstone of the post-RSG evolution at warm phases. The star is exceptionally known for showing a dramatic outburst in 2000$-$2001 (the ``Millenium outburst'') that led to a decrease in the effective temperature by 3\,000 K due to the ejection of a hydrogen shell of 0.03 M$_\odot$ \citep{lob03}. The strongest observed outburst of the star, however, occurred in the mid 40s and resulted in a deeper drop in the visual magnitude with a much longer duration compared to the event of 2000 \citep{Marav21}. The latest outburst of $\rho$ Cas was recorded in 2013 and led to a drop in temperature similar to that of the Millenium outburst \citep{Kraus19}. Beyond the Milky Way, it is surprising that only Var A in M33 has been assigned a solid post-RSG status resulting from a remarkable variability in the spectral type \citep{1987AJ.....94..315H,2006AJ....131.2105H}. Specifically, in the 1950s the star was observed as a high-luminous F-type star that, following a rapid photometric decline of $\sim3$ mag, changed its type to M-type supergiant in late 1980s. Currently, the star has returned to its pre-outburst warm state. The high energetic event with a duration of 45 years is attributed to the firm establishment of a pseudo-photosphere with yet questionable such long-term maintenance. The heavily obscured star Object X in M33 is proposed to be an analogue to Var A, undergoing multiple ejections of mass over several decades \citep{2011ApJ...732...43K}. Also in M33, \cite{2017A&A...601A..76K} suggested the discovery of a YHG, being the second extragalactic star to display a photometric event indicative of a YHG outburst that is paired with pulsational activity and changes in the color.

In principle, the limited number of identified YHGs is ascribed to the brief duration of the post-RSG phase. However, YHGs also shed substantial amounts of gas, which can lead to a self-obscuration by the dust formed out of the cooled ejecta. It is therefore expected that the eruptive nature of YHGs can potentially render several counterparts inaccessible. From the point of view of an observer, the identification of a YHG suggests a collective process that involves the precise determination of the stellar parameters, the investigation of the atmospheric dynamics, and importantly, the long-term monitoring of the target for capturing the episodes of enhanced mass loss. Additional to these prerequisites, moderate to high resolution spectroscopy is needed to expose the large intrinsic broadening of the lines that is caused by the large-scale atmospheric motion.

Being intrigued by the lack of identified YHGs and $\rho$ Cas analogues in nearby galaxies, we aim to investigate the evolutionary properties of luminous evolved stars in the Large and Small Magellanic Cloud (LMC and SMC, respectively). We selected stars from \cite{deJag98} that are historically classified as hypergiants by means of their high luminosity and broad emission in H$\alpha$, but lack a systematic study that targets the aforementioned diagnostics of a YHG. We do not limit ourselves to the FG spectral types but extend to the earlier ones, in search of post-RSGs that crossed the YV and currently appear as counterparts of other distinct classes (e.g. low-luminosity LBVs). The paper is structured as follows: In Section \ref{data}, we present our sample, the echelle spectroscopy as well as the archival multi-band and time-series photometry. In Section \ref{tempr}, we measure spectroscopic temperatures by means of a calibration scheme that is based on synthetic models of low-gravity stars. To derive luminosities, we assemble and fit the spectral energy distributions (SEDs) of the stars in Section \ref{seds}. In Section \ref{asas_per}, we present the $V$-band time series of the stars and extract the peak frequencies from their multi-periodic content. An insight into the dynamics of the ejected gas and of the upper atmosphere is given in Section \ref{kinema}. We identify the hypergiant HD269953 as a star that is orbited by a disk-hosting companion in Section \ref{hd269953}. We discuss on the evolutionary state of the stars in Section \ref{discuss}, before summarizing the findings of the study in Section \ref{summary}.

\section{The Data}
\label{data}

The review study by \cite{deJag98} contains a compilation of the most luminous observed stars in the Galaxy and other galaxies from the up to that date literature. The catalogue essentially includes identified LBVs, supergiants, and stars classified as hypergiants (Ia+) mostly of A and later type. In the present study, we update the properties and explore the circumstellar environment of eight evolved hypergiants in the LMC and SMC from the above catalogue. Of the selected stars, HD268757 is classified as a supergiant that was reported to display a shift in the spectral type from G5 to G8 \citep{1979A&AS...38..151V}. The objects are listed in Table \ref{sample}, along with their coordinates from SIMBAD, host galaxy, spectral type as in \cite{deJag98}, and the alternative designations in the literature.

\begin{table*}      % is used to refer this table in the text
\caption{\label{sample} List of studied stars.}
\begin{tabular}{l c c c c | l} 
\hline\
Object ID & RA & DEC & Galaxy & Spectral type$^a$ & Alternative IDs \\
\hline
HD268757 & 04:54:14.3 & $-$69:12:36.4 & LMC & G8 Ia & R59, Sk$-$69 30\\
HD271182 & 05:21:01.7 & $-$65:48:02.4 & LMC & F8 Ia$^{+?}$ & R92, Sk$-$65 48\\
HD269723 & 05:32:25.0 & $-$67:41:53.6 & LMC & G4 Ia$^+$ & R117, Sk$-$67 178\\
HD269953 & 05:40:12.2 & $-$69:40:04.9 & LMC & G0 Ia$^+$ & R150, Sk$-$69 260\\
HD271192 & 05:21:41.0 & $-$65:52:07.2 & LMC & A0 Ia$^+$ & R98, Sk$-$65 53\\
HD270086 & 05:45:16.6 & $-$68:59:52.0 & LMC & A1 Ia$^+$ & R153, Sk$-$69 299, S143\\
HD33579	& 05:05:55.5 & $-$67:53:10.9 & LMC & A3 Ia$^+$ & R76, Sk$-$67 44, S171\\
HD7583	& 01:13:30.5 & $-$73:20:10.3 & SMC & A0 Ia$^+$ & R45, S57\\
\hline
\footnotesize{$^a$as in \cite{deJag98}}
\end{tabular}
\end{table*}

\subsection{Echelle spectroscopy}
\label{echelle}

Observations of the targets were taken with the Fiber-fed Extended Range Optical Spectrograph \citep[FEROS;][]{1999Msngr..95....8K} installed on the ESO-MPI telescope at La Silla, Chile. The spectrograph is equipped with a 2k$\times$4k EEV CCD detector and provides a resolving power R $\sim48\,000$ over the spectral range of 350$-$920 nm, which is spread over 39 echelle orders. The data were reduced and calibrated using the dedicated Midas pipeline for FEROS. In absence of observed telluric standard stars, we corrected the object spectra for telluric contamination using Molecfit \citep{2015A&A...576A..77S}, a software package which generates a synthetic transmission model based on the atmospheric profile during the observations. For easier cosmic ray removal, the observations of each object were split into two consecutive exposures, which were then averaged to increase the signal to noise (S/N). The log of observations with FEROS is given in Table \ref{obslog}, which lists the star identification, $V$-band magnitude from \cite{Hog00}, observation date followed by the modified julian date, exposure time, value of airmass, and the S/N of the averaged spectrum. We normalized the spectra by fitting the stellar continua with a group of low-order splines. As a final step, we corrected the data for the observed radial velocity by cross correlating against low-gravity model spectra. The effective temperature of the templates was initially chosen to be in agreement with the spectral type of the stars, and the process was repeated upon derivation of the accurate temperatures (see Section \ref{tempr}).

\begin{table*}
\centering  
\caption{\label{obslog} Log of FEROS observations.}
\begin{tabular}{l c c c c c c}
\hline
Object ID & V (mag) & Date & MJD & Exp. Time (s) & Airmass & S/N \\
\hline
HD268757 & 10.5 & 2017-08-27 & 57992.3376 & 2$\times$400 & 1.48 & 70 \\
HD271182 & 9.7 & 2016-12-07 & 57729.1445 & 2$\times$400 & 1.28 & 140 \\
         &     & 2017-08-28 & 57993.3092 & 2$\times$400 & 1.60 & 100 \\
HD269723 & 10.1 & 2014-11-28 & 56989.2134 & 2$\times$430 & 1.28 & 110 \\
         &     & 2016-10-31 & 57692.2234 & 2$\times$400 & 1.36 & 80 \\
         &     & 2017-08-27 & 57992.3501 & 2$\times$400 & 1.51 & 80 \\
HD269953 & 10.1 & 2014-11-28 & 56989.2394 & 2$\times$430 & 1.31 & 110 \\
         &     & 2015-10-17 & 57312.3032 & 2$\times$400 & 1.34 & 100 \\
         &     & 2016-07-29 & 57598.4369 & 2$\times$500 & 1.51 & 130 \\
         &     & 2016-10-31 & 57692.2503 & 2$\times$400 & 1.36 & 55 \\
         &     & 2017-08-25 & 57990.3736 & 2$\times$400 & 1.48 & 50 \\
HD271192 & 10.6& 2016-12-07 & 57729.1561 & 2$\times$900 & 1.27 & 140 \\
         &     & 2017-08-28 & 57993.3209 & 2$\times$900 & 1.53 & 85 \\
HD270086 & 10.2& 2017-08-27 & 57992.3627 & 2$\times$800 & 1.50 &  90 \\
HD33579	 & 9.1 & 2017-08-28 & 57993.2868 & 2$\times$400 & 1.68 & 110 \\
HD7583	 & 10.2& 2017-08-27 & 57992.2520 & 2$\times$900 & 1.41 & 100 \\
\hline
\end{tabular}
\end{table*}

\begin{figure*}
\centering
\caption{\label{overview} Selected portions of the FEROS spectra highlighting discussed features in this study. From bottom to top, the spectra are sorted by decreasing temperature, as this is inferred in Section \ref{tempr}.}
\includegraphics[width=15cm,clip]{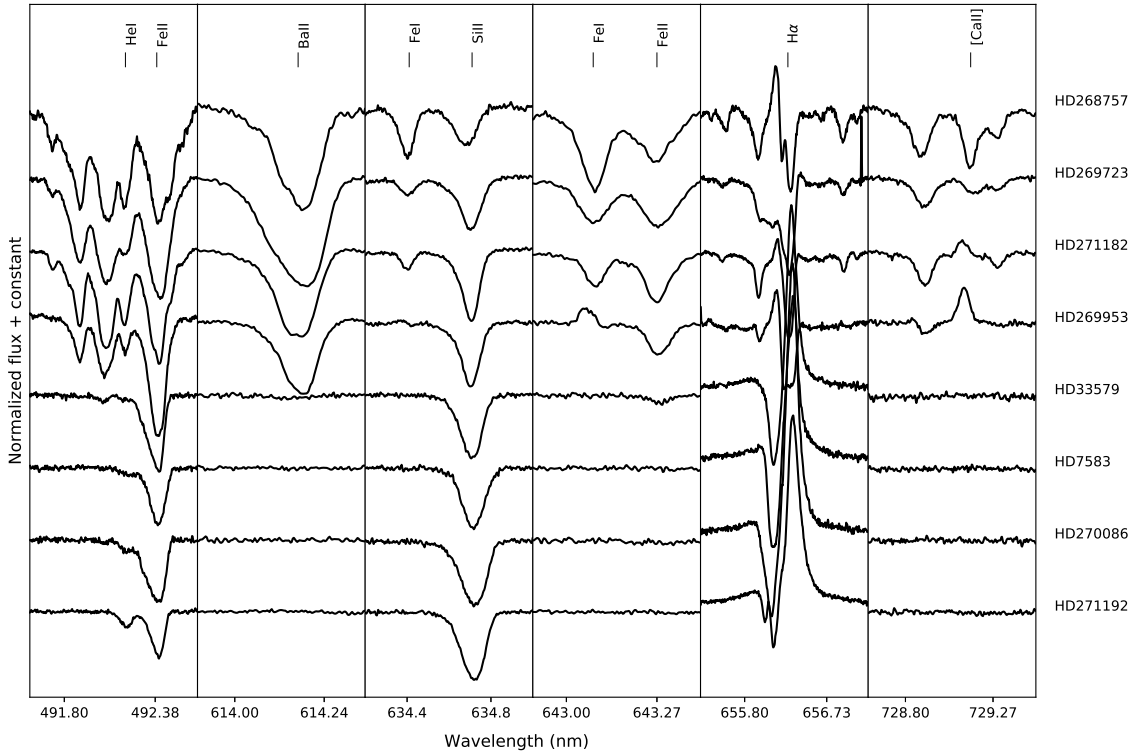}
\end{figure*}

In Fig. \ref{overview}, we display portions of the FEROS stellar spectra sorted by the temperature, as this is inferred in Section \ref{tempr}. The absence of neutral metal lines characterizes the four A-type stars, two of which show weak \ion{He}{I} that signifies the transition in the spectral type from late B to early A. The A-type stars in common display P Cygni in H$\alpha$, which is typical of the class of supergiants at our measured temperatures (Section \ref{ha_early}). On shifting to the FG types, H$\alpha$ is filled with emission components that trace the shocked ejecta (Section \ref{kinema_ha}). In addition, the low-excitation \ion{Ba}{II} line displays a strong shortward-displaced component that is frequently seen in YHGs and suggests circumstellar layers of deposited gas (Section \ref{kinema_ha}). Unsurprisingly, the lines of neutral metals strengthen with decreasing temperature, with the exception of HD269953 that shows emission in several of these lines and together with [\ion{Ca}{II}]. We attribute the latter findings to an evidence of a disk-hosting companion that orbits the hypergiant (Section \ref{hd269953}).    

\subsection{Archival multi-band photometry}
\label{photom}

Being amongst the visually brightest stars in the Clouds, our sample is included in the Tycho-2 Catalogue of bright stars \citep{Hog00} that were observed in the optical at the period $1990-1993$ with Hipparcos. The quality of data supersedes in quality the original Tycho-1 catalogue owing to improved reduction techniques and precision. Alongside Tycho we included sets of weighted mean photometry from the catalog of \cite{Merm87} compiled of individual measurements that were obtained through the $UBV$ filters from 1953 to 1985. The latter measurements point out a variable $V-$band magnitude for stars HD268757 and HD271182 with an amplitude that reaches 0.23 and 0.17 mag, respectively. For the rest of our sample, $V-$band variability does not exceed 0.1 mag. The mean values are found to be in good agreement with those derived from the posterior Tycho-2 catalog, with the exception of HD268757 showing a discrepancy of $\sim0.3$ mag. The optical photometry is complemented with the recent \textit{Gaia} Early Data Release 3 \citep{Gaia20}, which is based on data collected between 2014 and 2017. For targets as bright as our stars, \textit{Gaia} provides a high precision photometry with uncertainty of $\sim0.3$ mmag in the G and $\sim1$ mmag in both G$_{BP}$ and G$_{RP}$.

The near-infrared energy distribution of the stars was assembled from the $JHK_{s}$ photometry from 2MASS \citep{Cutri03}. In the mid-infrared, our stars were identified as point sources in the \textit{Spitzer} SAGE photometric catalogue in the LMC \citep{Meix06} and SMC \citep{GordK11} with measurements at 3.6, 4.5, 5.8, and 8.0 $\mu$m. Infrared counterparts extending to longer wavelengths were retrieved from the AllWISE catalog \citep{Cutri14} at 3.4, 4.6, 12, and 22 $\mu$m (W1, W2, W3, and W4, respectively). Our sample was additionally cross matched against the infrared point source catalogue obtained with the AKARI satellite \citep{Kato12} yielding counterparts for hypergiants HD271182, HD269953, and HD271192, at 3.2, 7, 11, 15, and 24 $\mu$m (corresponding to the imaging filters N3, S7, S11, L15, and L24, respectively).

The eight luminous stars were cross-matched against the aforementioned catalogues using a search radius of 2\arcsec, which resulted in one counterpart per star. The photometry and uncertainties of the sample stars are listed in Table \ref{photometry}.

\subsection{Time-series photometry}
\label{asas3}

Time-series photometry in the $V$ band of our objects was obtained during the third phase of the All Sky Automated Survey \citep[ASAS;][]{2002AcA....52..397P}. The time span covered by ASAS$-$3 was approximately nine years starting from late 2000. We accessed the data through the query interface of the survey\footnote{http://www.astrouw.edu.pl/asas/?page=aasc\&catsrc=asas3}, identifying one photometric source per target star using the resolution of 15\arcsec of the ASAS catalogue. The data files contain photometry obtained with five aperture diameters for reducing the background noise in fainter stars and avoiding blending effects in crowded ones. Given the $V$-band range  $9-10.5$ mag of our targets, we proceeded with the moderate aperture photometry (designated as MAG\_3). A comparison with the data obtained with the other apertures showed insignificant differences, with the exception of stars HD269723 and HD269953 the lightcurves of which were offset by less than 0.1 mag when observed with the largest aperture. With respect to the quality of the data, we selected datapoints that were assigned grades A and B. The nine-year optical photometry is presented in Section \ref{asas_per}, where it is followed by a Fourier analysis in search of periodic signals.

\begin{table*}
\caption{Multi-band photometry of the studied stars in the Magellanic Clouds. Uncertainties are given in parentheses.}\label{photometry}
\centering 
\begin{tabular}{l | c c c c c c c c} 
\hline
\multirow{2}{*}{Object ID}& $U$ & $B$ & $V$ & $B_{T}$ & $V_{T}$ & G$_{BP}$ & G & G$_{RP}$ \\ 
& (mag) & (mag) & (mag) & (mag) & (mag) & (mag) & (mag) & (mag) \\ 
\hline
HD268757 & 13.04 (0.22) & 11.71 (0.09) & 10.18 (0.09) & 12.08 (0.13) & 10.49 (0.05) & 10.52 (0.01) &  9.80 ($<0.01$) &  9.00 (0.01)\\ 
HD271182 & 10.74 (0.07) & 10.29 (0.05) &  9.70 (0.03) & 10.53 (0.04) &  9.70 (0.03) &  9.78 ($<0.01$) &  9.44 ($<0.01$) &  8.93 ($<0.01$)\\ 
HD269723 &              &              &              & 11.52 (0.08) & 10.12 (0.04) & 10.26 (0.01) &  9.71 ($<0.01$) &  9.01 (0.01)\\ 
HD269953 & 11.45 (0.03) & 10.83 (0.03) &  9.95 (0.01) & 11.04 (0.06) & 10.05 (0.03) & 10.13 ($<0.01$) &  9.76 ($<0.01$)&  9.21 ($<0.01$)\\ HD271192 & 10.24 (0.20) & 10.68 (0.03) & 10.56 (0.03) & 10.73 (0.04) & 10.59 (0.05) & 10.63 ($<0.01$) & 10.55 ($<0.01$) & 10.38 ($<0.01$)\\ 
HD270086 & 10.26        & 10.52 (0.02) & 10.28 (0.02) & 10.51 (0.04) & 10.22 (0.04) & 10.36 (0.01) & 10.23 ($<0.01$) &  9.97 (0.01)\\ 
HD33579	 &  9.07 (0.03) & 9.32 (0.02) &  9.13 (0.02) &  9.38 (0.02) &  9.13 (0.02) &  9.20 (0.01) &  9.09 ($<0.01$) &  8.86 (0.01)\\ 
HD7583	 & 10.00 (0.02) & 10.35 (0.02) & 10.21 (0.02) & 10.36 (0.03) & 10.20 (0.03) & 10.19 ($<0.01$) & 10.11 ($<0.01$) &  9.92 (0.01)\\ 
\hline\hline
\multirow{2}{*}{Object ID}& \multicolumn{3}{c}{$J$} & \multicolumn{2}{c}{$H$} & \multicolumn{3}{c}{$K_{s}$}\\
& \multicolumn{3}{c}{(mag)} & \multicolumn{2}{c}{(mag)} & \multicolumn{3}{c}{(mag)}\\
\hline
HD268757 &  \multicolumn{3}{c}{8.02 (0.02)} & \multicolumn{2}{c}{7.64 (0.03)} &  \multicolumn{3}{c}{7.45 (0.02)}\\ 
HD271182 &  \multicolumn{3}{c}{8.48 (0.03)} & \multicolumn{2}{c}{8.32 (0.05)} &  \multicolumn{3}{c}{8.22 (0.03)}\\ 
HD269723 &  \multicolumn{3}{c}{8.23 (0.03)} & \multicolumn{2}{c}{7.88 (0.03)} &  \multicolumn{3}{c}{7.69 (0.03)}\\ 
HD269953 &  \multicolumn{3}{c}{8.59 (0.03)} & \multicolumn{2}{c}{8.33 (0.05)} &  \multicolumn{3}{c}{8.02 (0.02)}\\ 
HD271192 & \multicolumn{3}{c}{10.20 (0.03)} & \multicolumn{2}{c}{10.15 (0.02)} & \multicolumn{3}{c}{10.06 (0.02)}\\ 
HD270086 & \multicolumn{3}{c}{9.69 (0.02)} & \multicolumn{2}{c}{9.62 (0.02)} &  \multicolumn{3}{c}{9.55 (0.02)}\\ 
HD33579	 & \multicolumn{3}{c}{8.69 (0.02)} & \multicolumn{2}{c}{8.59 (0.04)} &  \multicolumn{3}{c}{8.50 (0.02)}\\ 
HD7583	 & \multicolumn{3}{c}{9.80 (0.02)} & \multicolumn{2}{c}{9.70 (0.02)} &  \multicolumn{3}{c}{9.62 (0.02)}\\ 
\hline\hline
\multirow{2}{*}{Object ID}& [3.6] & [4.5] & [5.8] & [8] & $W_{1}$ & $W_{2}$ & $W_{3}$ & $W_{4}$ \\
& (mag) & (mag) & (mag) & (mag) & (mag) & (mag) & (mag) & (mag) \\
\hline
HD268757 & 7.20 (0.03) &  7.55 (0.02) &  7.25 (0.02) &  7.29 (0.03) &  7.07 (0.03) &  7.09 (0.02) &  6.80 (0.02) & 5.39 (0.03) \\ 
HD271182 & 8.02 (0.02) &  7.94 (0.02) &  7.92 (0.02) &  7.88 (0.02) &  8.12 (0.02) &  8.05 (0.02) &  7.88 (0.02) & 7.72 (0.08)  \\ 
HD269723 & 7.45 (0.03) &  7.31 (0.02) &  7.25 (0.02) &  7.24 (0.02) &  7.34 (0.04) &  7.28 (0.02) &  6.21 (0.02) & 1.58 (0.02)  \\ 
HD269953 & 7.28 (0.03) &  6.70 (0.02) &  6.23 (0.03) &  5.16 (0.03) &  7.27 (0.04) &  6.63 (0.02) &  4.06 (0.01) & 1.21 (0.02) \\ 
HD271192 & 9.92 (0.02) &  9.85 (0.02) &  9.77 (0.03) &  9.72 (0.02) &  9.90 (0.02) &  9.84 (0.02) &  9.65 (0.03) & 9.19  \\ 
HD270086 & 9.32 (0.04) &  9.26 (0.02) &  9.20 (0.02) &  9.17 (0.03) &  9.37 (0.02) &  9.30 (0.02) &  9.06 (0.03) & 8.67 (0.25) \\ 
HD33579	 & 8.35 (0.03) &  8.28 (0.03) &  8.23 (0.02) &  8.18 (0.02) &  8.35 (0.02) &  8.29 (0.02) &  8.09 (0.02) & 7.95 (0.09) \\ 
HD7583	 & 9.48 (0.03) &  9.43 (0.01) &  9.38 (0.01) &  9.34 (0.01) &  9.52 (0.02) &  9.50 (0.02) & 9.32 (0.03) & 8.86 \\ 
\hline\hline
\multirow{2}{*}{Object ID}& \multicolumn{2}{c}{N3} & S7 & \multicolumn{2}{c}{S11} & L15 & \multicolumn{2}{c}{L24}\\
& \multicolumn{2}{c}{(mag)} & (mag) & \multicolumn{2}{c}{(mag)} & (mag) & \multicolumn{2}{c}{(mag)}\\
\hline
HD271182 & \multicolumn{2}{c}{8.10 (0.04)} & 7.88 (0.01) & \multicolumn{2}{c}{7.71 (0.02)} &      &    \\ 
HD269953 & \multicolumn{2}{c}{7.39 (0.05)} & 5.73 ($<0.01$)& \multicolumn{2}{c}{3.84 (0.09)} & 3.24 (0.07) & \multicolumn{2}{c}{1.99 (0.02)}\\ 
HD271192 & \multicolumn{2}{c}{9.94 (0.04)} & 9.68 (0.04) & \multicolumn{2}{c}{9.63 (0.06)} & 9.52 (0.08) & \multicolumn{2}{c}{9.34 (0.27)}\\
\hline
\end{tabular}
\end{table*}

\section{Temperature determination}
\label{tempr}

Determining precise stellar parameters is essential for identifying the evolutionary channels that drive the stars. In turn, a link can be drawn between the evolutionary properties and the dynamics that shape the circumstellar environment. The spectral types of the eight stars from the literature (Table \ref{sample}) were derived from spectroscopy that dates back several decades upon comparison to classification standards \cite[][and references within]{deJag98}. More recently, the temperatures and luminosities of part of our FG sample were revised by \cite{2012ApJ...749..177N} by conversion of their $J-K$ color and assuming a constant value for the reddening. The method, however, can be sensitive to near-infrared emission from the hot dust, winds, and disks. Importantly and as stressed in \cite{2019A&A...631A..48V}, YHGs are prone to changes in the opacity of their outer layers, which results in colors reddened under a poorly understood absorption law. In the current section, we proceed to derive spectroscopic temperatures from the analysis of the FEROS data. 

A robust method for determining the temperatures of warm stars is by calibrating the ratios of the depths of metal absorption lines in the normalized spectra \citep{2000A&A...358..587K,1994PASP..106.1248G}. The strong advantage of the method against using the absolute depths is that it can effectively cancel out the broadening effects due to stellar rotation and the macroturbulence ($v_{\text{mac}}$), as both mechanisms only contribute in reshaping the lines assuming these are devoid of blends \citep{1991PASP..103..439G}. Selecting lines that are low on the curve of growth can further alleviate the dependence on the microturbulent velocity ($v_{\text{t}}$) \citep{1993KFNT....9...27S}. Ideally, lines of ionized metals should be avoided as these are particularly sensitive to the surface gravity. 

Calibration schemes for yellow supergiants such as that of \cite{2000A&A...358..587K}, are built over Galactic star samples and are not suitable for the proper classification of our FG-type stars in the LMC. Many of the lines from the above scheme were found to be rather weak in our spectra, whereas measuring the depth of others was undermined by the uncertainty in the level of local continuum due to wing blends. To proceed with the discussed method, it was necessary to employ new temperature-sensitive sets of lines that would be applicable to our data. As the limitations described constrain significantly the number of lines suitable for classification, we included in our analysis lines of ionized metals and fixed the surface gravity of the models to the lowest value available.

For the purpose of calibrating temperature using the depth ratios of atmospheric lines, we generated synthetic spectra using iSpec, a software framework that includes a collection of python-wrapped radiative transfer and spectral synthesis codes \citep{Blanco14}. To characterize our FG sample, we generated a grid of spectra using the synthesis code Turbospectrum \citep{Plez12} and MARCS model atmospheres \citep{Gustaf08}. The models were calculated at $\log g=0.0$ dex assuming spherical geometry and mixing length parameter $\ell/H=1.5$, and cover temperatures from 2\,500 K to 8\,000 K. For classifying the hotter stars, spectra were synthesized using SPECTRUM radiative transfer code \citep{Grey94} and ATLAS9 plane-parallel model atmospheres \citep{Castelli03}. The latter models assume $\ell/H=1.25$. We adopted the lowest available value of $\log g=2.0$ dex for temperatures extending up to $T_{\text{eff}} = 14\,000$ K. In both codes, calculations were performed in local thermodynamic equilibrium (LTE). The atomic lines chosen for spectral synthesis were extracted from the Vienna Atomic Line Database \citep[VALD;][]{kupka11}.

The solar abundances were taken from \cite{Grev07} and were scaled to the suitable metallicity. In principle, the metallicity of the models was chosen to match that of the host, which for our stars in the LMC was set to the value [M/H] $=-0.5$ \citep{2006AJ....132.1630G}. As the lines of HD33579 and HD270086 were shown to be systematically stronger than those of the theoretical models, we complemented the analysis with models at solar metallicity. For HD33579, abundances are known to indeed approach the values of the Galactic $\alpha$ Cygni \citep{1968MNRAS.139..313P}. Respectively, the lines of HD7583 in the SMC were stronger than those of ATLAS9 models at [M/H] $=-1.0$ \citep{2009AJ....138..517P}, regardless of the choice of $v_{\text{t}}$. The abnormal high metal abundances of the star were also reported by \cite{Wolf73}. The classification of HD7583 was carried out over the same scheme of line-depth ratios as for the LMC stars.

We tested the sensitivity of the calibration scheme on the microturbulence. The analysis was carried out using different values of $v_{\text{t}}$ that are consistent to the literature studies and, at the same time, provide a reasonable fit to our data. Accordingly, for the ATLAS9 models we used a $v_{\text{t}}$ of 6 and 8 km~s$^{-1}$, which agrees with the observations of early-A type supergiants \citep{1995ApJS...99..659V,2000ApJ...541..610V}. For the MARCS models, we proceeded with $v_{\text{t}}$ at 10 and 12 km~s$^{-1}$, which we found to be essential for fitting the strong line cores of the FG sample. Consistent supersonic values of $v_{\text{t}}$ are measured for the few known YHGs \citep[see][]{deJag98} and are attributed to the motion of stochastically distributed shocked layers that can not be spectroscopically distinguished from the classical small-scale turbulent motion \citep{deJag98,Lob98}. 

With grids of synthetic models in hand, we run a process for assessing the model temperature as function of the depth ratio of neighbor lines. To minimize errors introduced by possible continuum displacements in the observations, the separation threshold between the adjacent lines was set to 0.2 nm for the MARCS spectra. Due to the limited number of lines, this value was increased to 0.4 nm for the hotter ATLAS9 models. On proceeding with the cooler grid, we discarded lines shortward  of 530 nm to avoid heavily blended lines within the iron forest. Upon comparison to the observed spectra, we discarded line sets the measurement of which was restricted by the noise and normalization process. In addition, several lines of the A-type stars are shown to display asymmetries caused by velocity gradients in the upper atmosphere. We therefore excluded lines with the strongest such effect, namely the \ion{Fe}{II} lines in multiplets 27, 38, 49, and particularly, 42. We selected three and four line pairs for the temperature calibration of A and FG luminous stars, respectively. The wavelengths and identifications of these lines are listed in Table \ref{coeff}.

We evaluated the sensitivity of the depth ratios to the line broadening to ensure that the efficiency of the method is not undermined by blends. For each of the two grids of models, we broadened lines using two sets of $v$sin$i$ and $v_{\text{mac}}$. For the A-type supergiants, the theoretical models of stellar evolution predict equatorial velocities of less than 10 km~s$^{-1}$ \citep{2012A&A...537A.146E, 2013A&A...558A.103G}. On the other hand, observations of BA-type supergiants in the Galaxy \citep{2012A&A...543A..80F} and of A-type supergiants in M31 \citep{2000ApJ...541..610V} verify $v$sin$i$ values up to roughly 30 km~s$^{-1}$. Bearing in mind that the relatively weaker winds of the stars in the Clouds would prevent losses of angular momentum, we accept the latter value as reasonable for our testing purposes. We pair this choice with $v_{\text{mac}}$ of 12 km~s$^{-1}$ in agreement to the value reported for $\alpha$ Cygni \citep{1988A&A...195..218B}. The second set of test parameters for the ATLAS9 grid neglects macroturbulence and assigns $v$sin$i$ a value of 15 km~s$^{-1}$ thereby suggesting an overall weaker line broadening. When proceeding with the FG stars, the two mechanisms exert opposite roles on the line shaping. Assuming negligible rotation as for a physically extended star, a fit to the FEROS data indicates $v_{\text{mac}}$ values that exceed $20-25$ km~s$^{-1}$. The large-scale atmospheric motion is the hallmark of late-type luminous stars that are found in a post-RSG \citep[e.g.][]{2016MNRAS.461.4071S} or post-AGB state \citep{2000A&A...359..138T}. Our first set of test parameters neglects so the rotational effect and assigns the MARCS spectra a strong macroturbulence of 30 km~s$^{-1}$. Alternatively and based on the observations of $\rho$ Cas, we applied a $v_{\text{mac}}$ of 20 km~s$^{-1}$ \citep{lob03} on top of a rotational broadening of 15 km~s$^{-1}$ that matches a line broadening of 25 km~s$^{-1}$ \citep{Lob98}.

\begin{figure}
\centering
\caption{\label{warm_calib} Calibration curves for inferring the temperatures of the early-A sample stars. The fit to the line-depth ratios of low-gravity ATLAS9 models at [M/H] = $-$0.5 is shown with solid and dotted lines, calculated for pairs of broadening values ($v$sin$i$, $v_{\text{mac}}$) = (30 km~s$^{-1}$, 12 km~s$^{-1}$) and (15 km~s$^{-1}$, 0), respectively. The models were generated with $v_{t}=8$ km~s$^{-1}$ (red) and 6 km~s$^{-1}$ (blue). The fit to the line-depth ratios of models at solar metallicity is shown with red dashed line, calculated with ($v$sin$i$, $v_{\text{mac}}$) = (30 km~s$^{-1}$, 12 km~s$^{-1}$) and $v_{t}=8$ km~s$^{-1}$.}
\includegraphics[width=5.5cm,clip]{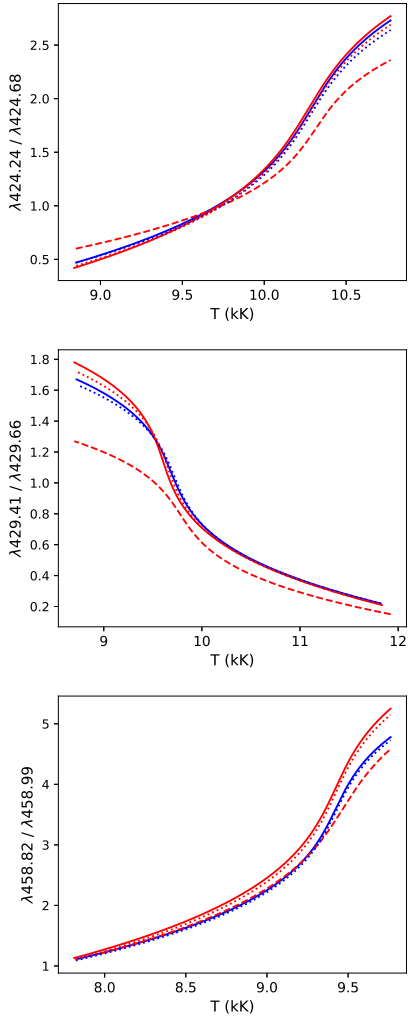}
\end{figure}

\begin{figure}
\caption{\label{cool_calib} Calibration curves for inferring the temperatures of our FG-type stars. We show the fit to the line-depth ratios of low-gravity MARCS models at [M/H] = $-$0.5 with solid and dotted lines, calculated for pairs of broadening values ($v$sin$i$, $v_{\text{mac}}$) = (1 km~s$^{-1}$, 30 km~s$^{-1}$) and (15 km~s$^{-1}$, 20 km~s$^{-1}$), respectively. The models were generated with $v_{t}=12$ km~s$^{-1}$ (red) and 10 km~s$^{-1}$ (blue).}
\includegraphics[width=8.6cm,clip]{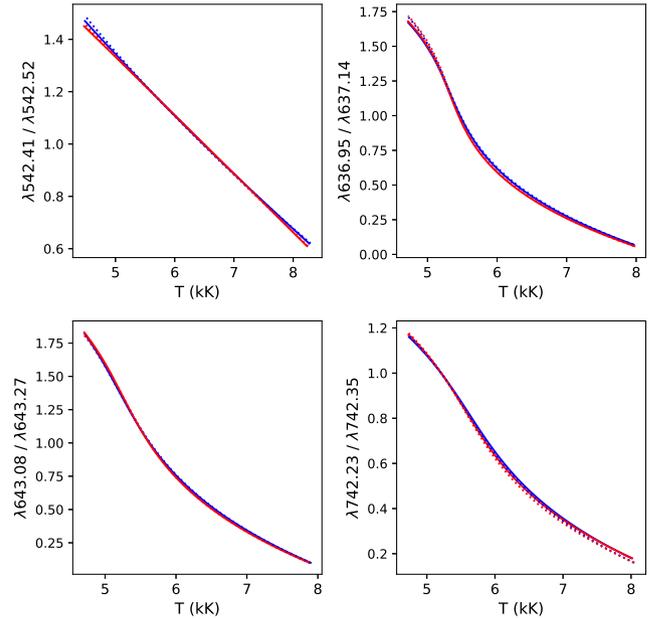}
\end{figure}

The line-depth ratios as function of the model temperature for the different broadening sets were fit by third-order polynomials (second-order for the ratio \ion{Fe}{I} $\lambda542.41$ to \ion{Fe}{II} $\lambda542.52$), the coefficient of which are given in Table \ref{coeff}. The calibration curves are displayed in Fig. \ref{warm_calib} and \ref{cool_calib} for the classification of the A and FG luminous stars, respectively. For a fixed value of $v_{\text{t}}$, we find negligible or no dependence of the ratios on the two broadening parameters. The uncertainty of 2 km~s$^{-1}$ in the microturbulence of the ATLAS9 models yields discrepancies in the temperature of maximum 260 K. This disagreement is essentially due to microturbulence influencing differently the lines at different parts of the curve of growth by preserving or modifying the equivalent width. For the classification of our A stars, we adopt $v_{t}=8$ km~s$^{-1}$. The method is shown to be most sensitive to metallicity and particularly when proceeding with the ratio \ion{Ti}{II} $\lambda429.41$ to \ion{Fe}{II} $\lambda429.66$.

\begin{figure}
\centering
\caption{\label{warm_spectra} Fit of  the temperature-sensitive line pairs of the A-type stars (black) using ATLAS9 models (red).}
\includegraphics[width=8.4cm,clip]{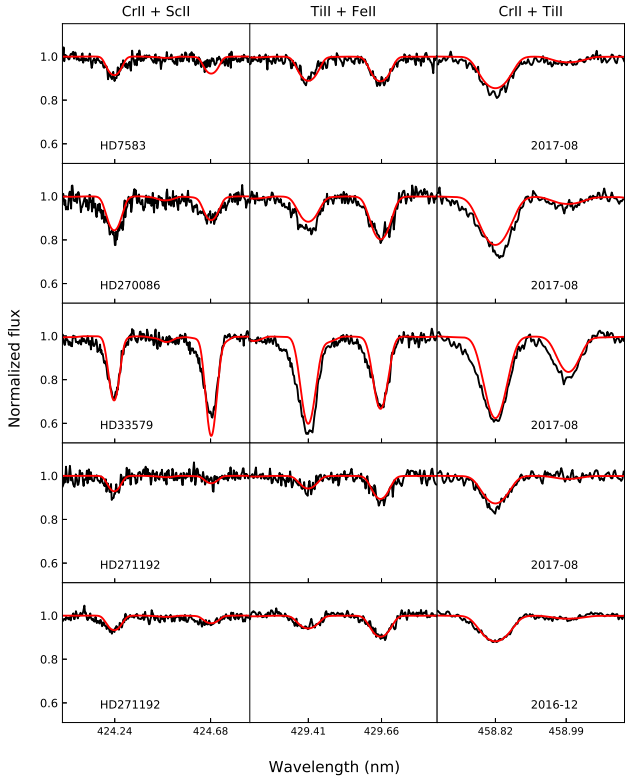}
\end{figure}

\begin{figure}
\caption{\label{cool_spectra} Same as in Fig. \ref{warm_spectra}, but for the spectroscopy of our four FG-type stars (black) and using MARCS models (red). Note the double emission profile of the \ion{Fe}{I} $\lambda643.1$ line of HD269953; the line ratio with the adjacent \ion{Fe}{II} was discarded during the temperature determination process.}
\includegraphics[width=8.6cm,clip]{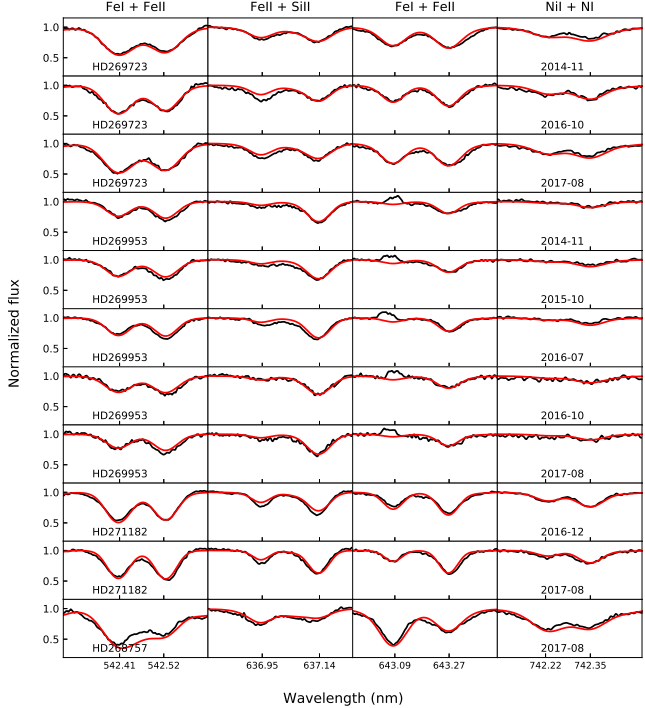}
\end{figure}

The depths of the observed lines were measured using IRAF task \textit{splot}. For each set of lines, we calculated the error in the ratio using the S/N of the local continuum. When more than one line set was measurable, the average value was derived by bootstraping across the different $T_{\text{eff}}$ measurements and their uncertainties. The resulting values, rounded to the nearest 50 K, are listed in Table \ref{final_param}. In Figs. \ref{warm_spectra} and \ref{cool_spectra}, we show the fit of the models with the resulting temperatures to the FEROS data of the A and FG stars, respectively. Having $T_{\text{eff}}$ fixed, the broadening parameters were determined by $\chi^2$ minimization. The shown ATLAS9 models in Fig. \ref{warm_spectra} have $v$sin$i$ values within 25$-$35 km~s$^{-1}$ for $v_{\text{mac}}$ in the range 12$-$15 km~s$^{-1}$. Definitely, the models can not reproduce the asymmetric lines caused by the velocity gradients in the atmospheres of the A-type supergiants. Moreover, discrepancies can be amplified due to departures from LTE \citep{2006A&A...445.1099P}. 
The MARCS synthetic spectra have $v_{\text{mac}}$ in the range 25$-$35 km~s$^{-1}$ when assuming negligible rotation. Figure \ref{cool_spectra} illustrates an appreciable fit of the data, with a variable and large broadening characterizing HD268757. We caution, however, that the lines of the latter star may be contaminated by a companion spectrum (see Section \ref{discuss}). The \ion{Fe}{I} $\lambda643.1$ emission of HD269953 is discussed in Section \ref{hd269953} and the corresponding line set was not used for the classification of the star.

\begin{table*}
\caption{Coefficients of polynomial $T_{\text{eff}}(r) = ar^3 +br^2 +cr + d$ for determining temperature (in kK) using synthetic line-depth ratio $r=R_{\lambda_{1}}/R_{\lambda_{2}}$ at [M/H] = $-$0.5. For characterizing the two metal-rich A-type stars, we provide values at solar metallicity shown in italic. The range of validity of the method is given in the last column.}         
\label{coeff}
\centering         
\begin{tabular}{c | c c c| c c | r r r r| c}
\hline
Model       & $v_{\text{t}}$ & $v$sin$i$ & $v_{\text{mac}}$ & $\lambda_1$ &  $\lambda_2$  &    \multirow{2}{*}{a}  &   \multirow{2}{*}{b}   &   \multirow{2}{*}{c}   &   \multirow{2}{*}{d} & $\Delta T_{\text{eff}}$\\
atmospheres & (km~s$^{-1}$)  & (km~s$^{-1}$) & (km~s$^{-1}$) & (nm)     &     (nm)     &       &       &       &  & (kK)\\
\hline
\multirow{4}{*}{MARCS} & \multirow{4}{*}{$10-12$} & \multirow{4}{*}{$1-15$} & \multirow{4}{*}{$30-20$} & \ion{Fe}{I} $\lambda542.41$ & \ion{Fe}{II} $\lambda542.52$ &   --     & $-$0.06 &  $-$4.36 & 10.91 & \multirow{4}{*}{$4.7-8$}\\
&&&& \ion{Fe}{II}~~$\lambda636.95$ & \ion{Si}{II}~~$\lambda637.14$ &  $-$1.38 &    4.78 &  $-$6.30 &  8.34 &\\
&&&& \ion{Fe}{I}~~$\lambda643.08$ & \ion{Fe}{II}~~$\lambda643.27$ &  $-$0.74 &    2.98 &  $-$4.97 &  8.35 &\\
&&&& \ion{Ni}{I}~~$\lambda742.23$ & \ion{Si}{I}~~$\lambda742.35$ &  $-$4.02 &   10.06 & $-$10.41 & 9.59 &\\ 
\hline
\multirow{6}{*}{ATLAS9} & \multirow{6}{*}{8} & \multirow{6}{*}{$15-30$} &\multirow{6}{*}{$0-12$} & \ion{Cr}{II}~~$\lambda424.24$ & \ion{Sc}{II}~~$\lambda424.68$ &     0.25 & $-$1.46 &     3.19 &  7.74 & $8.7-10.8$\\
&&&&&& \textit{0.68} & \textit{$-$3.53} & \textit{6.53} & \textit{6.06} & \\
&&&&\ion{Ti}{II}~~$\lambda429.41$ & \ion{Fe}{II}~~$\lambda429.66$ &  $-$2.18 &    7.46 &  $-$9.01 & 13.42 & $8.7-12$\\
&&&&&& \textit{$-$5.51} & \textit{13.15} & \textit{$-$11.48} & \textit{13.37}&\\
&&&&\ion{Cr}{II}~~$\lambda458.82$ & \ion{Ti}{II}~~$\lambda458.99$ &     0.05 & $-$0.60 &     2.53 &  5.64 & $7.7-9.8$\\
&&&&&& \textit{0.07} & \textit{$-$0.72} & \textit{2.85} & \textit{5.48} &\\
\hline 
\end{tabular}
\end{table*}

\begin{table}
    \caption{\label{final_param} Inferred parameters for the eight stars. Numbers in parentheses list 2$\sigma$ error estimates.}
    \centering
    \begin{tabular}{lcc|c|c}
    \hline
    Star & Date & $T_{\text{eff}}$ & $\log(L/L_{\odot})$ & $A_{V}$\\
    \hline 
    HD268757 & 2017-08 & 4\,950 (150) & 5.64 (0.03) & 0.92 (0.09) \\
    HD271182 & 2016-12 & 6\,100 (50) & \multirow{2}{*}{5.65 (0.05)} & \multirow{2}{*}{0.66 (0.14)} \\
             & 2017-08 & 6\,500 (100) && \\
    HD269723 & 2014-11 & 5\,800 (100) & \multirow{3}{*}{5.76 (0.02)} & \multirow{3}{*}{1.31 (0.05)}\\
             & 2016-10 & 6\,000 (100) && \\
             & 2017-08 & 5\,800 (100) && \\
    HD269953 & 2014-11 & 7\,250 (100) & \multirow{5}{*}{5.79 (0.02)} & \multirow{5}{*}{1.28 (0.03)} \\
             & 2015-10 & 7\,050 (100) && \\
             & 2016-07 & 7\,100 (100) &&\\
             & 2016-10 & 7\,050 (250) &&\\
             & 2017-08 & 7\,300 (200) && \\
   \hline        
    HD271192 & 2016-12 & 10\,300 (250) &   \multirow{2}{*}{5.48 (0.02)} & \multirow{2}{*}{0.68 (0.01)} \\
             & 2017-08 & 10\,400 (400) && \\
    HD270086 & 2017-08 & 10\,150 (250) & 5.68 (0.02) & 0.94 (0.01)\\
    HD7583   & 2017-08 &  9\,800 (150) & 5.77 (0.02) & 0.69 (0.01) \\
    HD33579  & 2017-08 &  9\,000 (50) & 5.94 (0.02) & 0.69 (0.01)\\
    \hline
    \end{tabular}
\end{table}

\section{Spectral Energy Distributions}
\label{seds}

The wealth of archival photometric data spanning the optical to the mid-infrared region are employed to assemble the SEDs of the stars. Post-RSGs are often enshrouded by dust, especially if they have recently departed from the RSG phase when they shed most of their mass. However, the eruptions during the YHG phase may also be responsible for dust formation \citep{1990ApJ...351..583J,2011A&A...534L..10L,2020A&A...635A.183K}. By fitting the SEDs with photospheric models, we determined the luminosities and refine the positions of the stars on the evolutionary diagram.

The flux $f_{\lambda}$ at a given wavelength $\lambda$, which is measured from a stellar object at a distance $d$ that is subject to extinction $A(\lambda)$, is given by
\begin{equation}
	f_{\lambda} = (R/d)^{2} \times F_{\lambda} \times 10^{-0.4A(\lambda)}
\end{equation} 
where $R$ is the stellar radius and $F_{\lambda}$ is the surface flux. The photometric data were fit using grids of ATLAS9 photospheric models from \cite{2011MNRAS.413.1515H} with the lowest available values of $\log g$, and for the same chemical composition as that used for estimating the spectroscopic temperatures. We chose the set of models that were calculated with $\ell/H=1.25$, to match with the model atmospheres that were used in the spectral synthesis.

The distance to the LMC and SMC was set to the value $50.6\pm1.6$ kpc \citep{2011ApJ...729L...9B} and $62.1\pm1.9$ kpc \citep{2014ApJ...780...59G}, respectively. We converted between magnitudes and fluxes using the zero points and the effective wavelengths of the bandpass filters that are available at the SVO Filter Profile Service\footnote{http://svo2.cab.inta-csic.es/theory/fps/}. The photometry was fit by the models using a Levenberg–Marquardt minimization scheme, excluding the data points in the infrared that were in excess of the stellar continuum. The temperatures of the models was set to the averaged spectroscopic values from Table \ref{final_param}, and the synthetic data were reddened according to the extinction laws by \cite{1989ApJ...345..245C} using the standard value for $R_{V}=3.1$. We derived the stellar luminosities from the best-fit radii and list them, along with the resulting $A_{V}$ values, in Table \ref{final_param}. We assessed the uncertainties by bootstrapping over 1\,000 sets of input parameters, similar as in \cite{2015A&A...582A..42K}. 

\begin{figure}
\caption{\label{warm_seds} Spectral energy distribution of the A-type stars. Data are taken from \protect\cite{Merm87} (cyan), Tycho (magenta), \textit{Gaia} (blue), 2MASS (green), \textit{Spitzer} (black), WISE (red), and AKARI (pink). The black lines correspond to the best fitting photospheric models for the unreddened case (dotted) and upon extinction (solid).}
\includegraphics[width=8.7cm,clip]{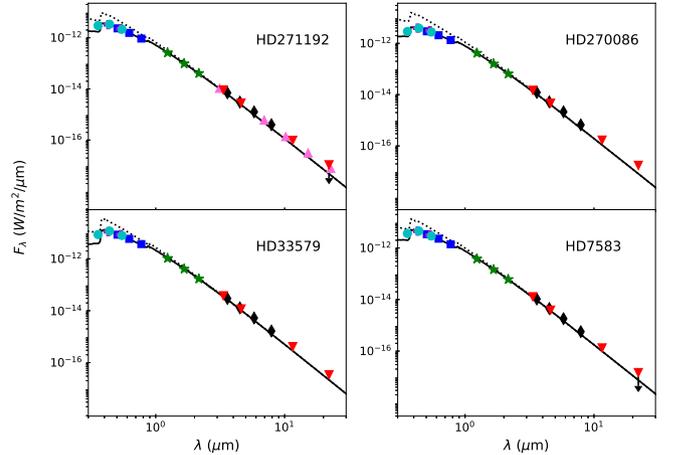}
\end{figure}

\begin{figure}
\caption{\label{cool_seds} Same as in Fig. \ref{warm_seds}, but for the FG-type stars. Note the infrared excess due to dust for HD269723 and, to a lesser extent, for HD268757. Dust along with wind excess likely emerging from the disk discussed in Section \ref{hd269953} is shown for HD269953.}
\includegraphics[width=8.7cm,clip]{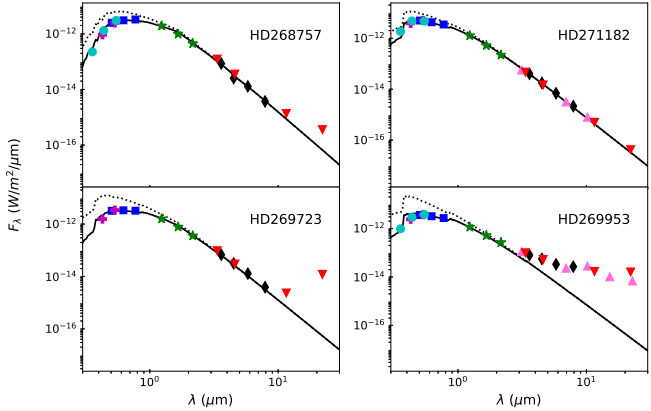}
\end{figure}

The best fitting models to the data are shown in Fig. \ref{warm_seds} for the A-type stars, and in Fig. \ref{cool_seds} for the FG-type stars. The weak offset between the synthetic data and the infrared observations for the A-type sample is attributed to the stellar winds the effect of which is not included in the model SEDs. The changes in stellar radii due to pulsations may also justify discrepancies between photometric sets that were obtained at different dates. Nevertheless and as mentioned in the following section, the time-series photometry in $V$, with the exception of HD33579, shows no detectable periodic modulation. Excess in the mid-infrared due to cool dust is prominent only within the FG sample, and especially for HD269723 and HD269953.

\section{Photometric variability}
\label{asas_per}

\begin{figure}
\centering
\caption{\label{lc} Time-series photometry in the $V$ band from ASAS$-$3.}
\includegraphics[width=8.7cm,clip]{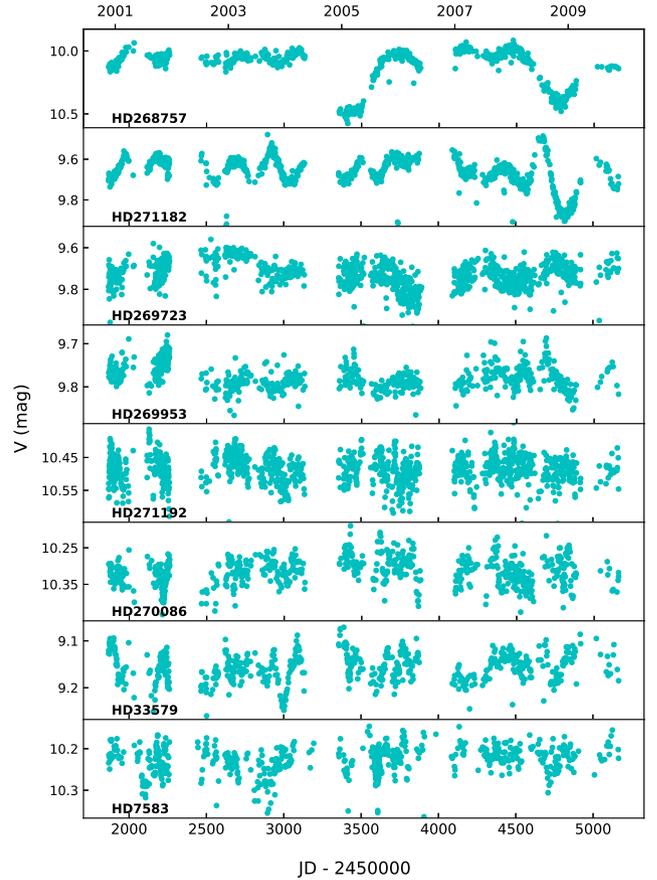}
\end{figure}

We present the $V-$band light curves of the eight stars from ASAS$-$3 in Fig. \ref{lc}. Outliers exceeding $\pm3\sigma$ from the mean magnitude were discarded and are not shown. An overview of the data highlights the variable baselines of HD268757 and HD271182. Also evident are the long-term (quasi-)periodic variability of HD269723 and the $\sim$100 day lasting drop in the brightness of HD33579 around JD2453000. The variability in the data set of the remaining objects does not exceed 0.2 mag.

We explored the multi-periodic content of the time series using \textsc{Period04} \citep{2005CoAst.146...53L}. The software employs a discrete Fourier transform algorithm to construct the amplitude spectrum and extract the peak frequency. The residuals of the sinusoidal fit are then subject to a new Fourier analysis. At each step of the prewhitening sequence, the extracted frequencies, amplitudes, and oscillation phases are adjusted using least-square minimization. The iterative process comes typically to its end when the power spectrum no longer contains peak frequencies of significance. 

The bandwidth for the detection of peaks was determined by the Nyquist frequency, which had the value 0.25 c d$^{-1}$ for four of the eight stars (HD268757, HD269953, HD270086, and HD33579). Due to denser sampling pattern, this upper frequency limit was found to be 26 c d$^{-1}$ for HD271182, and 15 c d$^{-1}$ for the remaining objects. Nevertheless, our inspection of the extended range failed to recover significant peaks other than the alias frequency at 1 c d$^{-1}$ and its harmonics. We hence constrained the detection of peaks to 0.25 c d$^{-1}$ for all stars. Of the retrieved frequencies per star, we discarded those that were not resolved from already identified neighboring peaks using the Rayleigh frequency resolution. The latter value is defined as 1$/$T, where T is the data set length, namely 3\,200 d. Exceptionally, the time span of HD268757 and HD271182 was truncated to JD2453100 and JD2454500, respectively, dates prior to the observed high-amplitude drops in the magnitude. Harmonic frequencies were also identified using the Rayleigh resolution. Lastly, we considered reliable frequencies those being higher than 2$/$T c d$^{-1}$.

\begin{figure}
\centering
\caption{\label{fourier} Power spectra of the ASAS$-$3 time series (black) and of the pre-whitened data at the end of the Fourier analysis (red). We mark the identified peak frequencies and the threshold 2/T (dashed line) below which frequencies are not considered reliable.}
\includegraphics[width=8.8cm,clip]{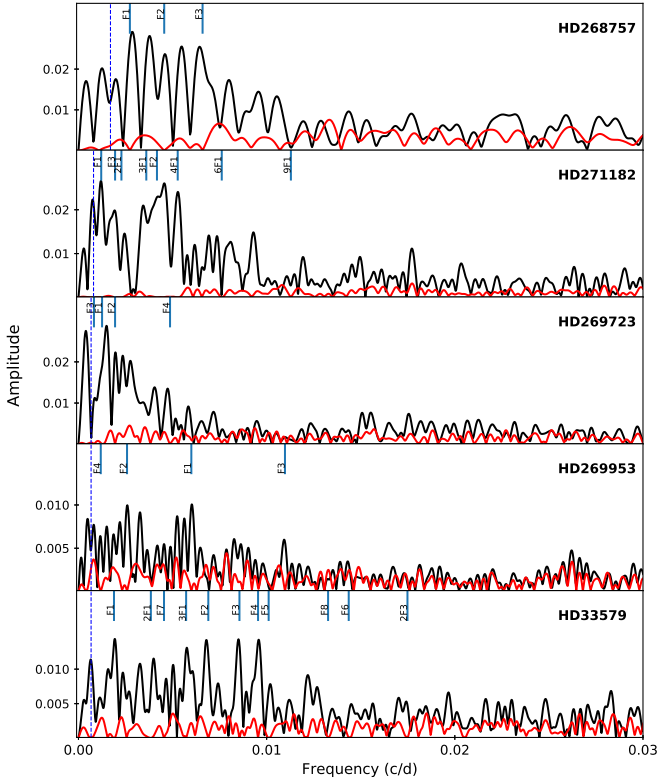}
\end{figure}

We extracted and report the peak frequencies for five of the objects in Table \ref{fourier_tab}. Non-listed stars were found to display noisy Fourier spectra, with peaks that were either not statistically significant or failed to minimize the residuals of the sinusoidal fit. A by eye inspection of the fit quality could further verify these frequencies as non-credible. For each of the listed stars, we display in Fig. \ref{fourier} the first Fourier spectrum alongside the noise spectrum at the end of the pre-whitening process. Apparently, the strongest signal is detected for our coolest stars HD268757, HD271182, HD269723, and the most luminous of our objects, HD33579.

\begin{table}
    \caption{\label{fourier_tab} Peak frequencies and harmonics identified during the pre-whitening of the ASAS$-$3 data.}
    \centering
    \begin{tabular}{lcccc}
    \hline
    Star & $\#$ & Frequency (c d$^{-1}$) & Amplitude (mag)\\
    \hline
    \multirow{3}{*}{HD268757} & F1 & 0.00273 &	 0.031\\
    & F2 & 0.00456 & 0.018\\ 
    & F3 &	0.00660 & 0.018\\
    \hline    
    \multirow{8}{*}{HD271182} & F1 & 0.00120 &  0.024\\ 
    & F2 & 0.00417 & 0.025\\ 
    & 3F1 &	0.00360 & 0.034\\ 
    & 2F1 & 0.00228 & 0.025\\ 
    & 4F1 &	0.00527 & 0.027\\ 
    & 6F1 &	0.00761 & 0.007\\ 
    & 9F1 &	0.01129 & 0.004\\ 
    & F3 &	0.00195 & 0.014\\    
    \hline
    \multirow{4}{*}{HD269723} & F1 & 0.00125 &	 0.028\\ 
    & F2 & 0.00194 & 0.031\\ 
    & F3 & 0.00082 & 0.020\\ 
    & F4 & 0.00487 & 0.009\\
    \hline
    \multirow{4}{*}{HD269953} & F1 & 0.00600 &	 0.008\\ 
    & F2 & 0.00258 & 0.010\\ 
    & F3 & 0.01097 & 0.005\\ 
    & F4 & 0.00118 & 0.009\\
    \hline
    \multirow{11}{*}{HD33579} & F1 & 0.00189 & 0.019\\ 
    & F2 & 0.00690 & 0.007\\ 
    & F3 & 0.00855 & 0.015\\ 
    & F4 & 0.00954 & 0.012\\ 
    & 2F1 &	0.00385 & 0.011\\ 
    & F5 & 0.01011 & 0.008\\ 
    & 2F3 &	0.01748 & 0.007\\ 
    & 3F1 &	0.00572 & 0.011\\ 
    & F6 & 0.01436 & 0.007\\ 
    & F7 & 0.00455 & 0.009\\ 
    & F8 & 0.01327 & 0.006\\
    \hline
    \end{tabular}
\end{table} 

\section{Kinematics of ejecta and upper atmosphere}
\label{kinema}

In the current section, we discuss findings on the dynamics of mass loss and gas ejecta. We focus mainly on the FG sample due to the wealth of information extracted from their active nature but also based on the availability of more than one spectrum per object for three of these, which allows insight on dynamical processes that take place in the atmosphere.

\subsection{H\texorpdfstring{$\alpha$}{} in the A-type stars}
\label{ha_early}

\begin{figure}
\centering
\caption{\label{warm_Ha} The H$\alpha$ line profile of the studied early-A type stars.}
\includegraphics[width=6cm,clip]{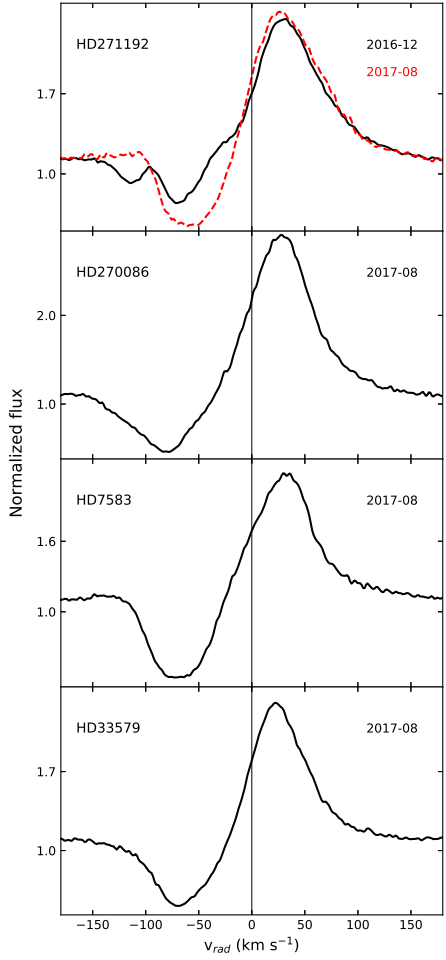}
\end{figure}

Our early-A sample displays an H$\alpha$ P Cygni profile with the emission being centered within $25-35$ km~s$^{-1}$ and the absorption core being found at velocities of $-70$ to $-90$ km~s$^{-1}$ (Fig. \ref{warm_Ha}). The strongest emission is shown for HD270086 at the same time that absorption extends shortward of $-100$ km~s$^{-1}$. The star strikes out of the four stars due to its high energetic wind output. The absorption component of HD271192 is shown to be variable over the different epochs. In specific, the 2016 data (black solid line in Fig. \ref{warm_Ha}) indicate outflow that extends to $-150$ km~s$^{-1}$ with superimposed emission that is centered at $-90$ km~s$^{-1}$. Within a year, the absorption line appears redshifted (red dashed line) and comparable in shape to that of HD7583. We interpret this variability as evidence of density inhomogeneities traveling through the wind. The emergence of such lines in the absorption components of H$\alpha$ is generally not uncommon for early-A supergiants \citep[e.g.][]{2012A&A...543A..80F}.

The observed H$\alpha$ components lie on weak Thomson scattering wings, which are formed in the wind and extend to $700-1\,000$ km~s$^{-1}$. Extended wings to a similar degree are observed in the H$\alpha$ P Cygni of warm hypergiants in M31 \citep{Humphr13}. However, the outflow velocities of the latter, measured from the absorption component of H$\alpha$, were found to be $3-5$ times larger than the respective values for the LMC and SMC stars. Moreover, the H$\alpha$ emission cores of the M31 stars were found to be several times stronger than those of our present sample. Importantly, the warm hypergiants of M31 additionally show P Cygni in \ion{Fe}{II} and \ion{Ca}{II} that, along with the observed excess in the near to mid-infrared due to winds, characterize a rather active mass-loss phase.

\subsection{Circumstellar environment of the late-type hypergiants}
\label{kinema_ha}

\cite{Sarg61} was one of the first studies to report the split of metal lines with $\chi_{\text{low}}<3$ eV in the spectrum of $\rho$ Cas. Asymmetric lines of metals such as \ion{Sr}{II}, \ion{Y}{II}, and \ion{Ba}{II} were suggested to comprise of two components; a shortward that grows with increasing excitation potential and emerges from circumstellar gas, and a longward core with a profile similar to that of the single lines. \cite{Lob98} argued that, the observed splitting is in fact a broad absorption line with the superposition of a static and narrow emission line of circumstellar origin. As the absorption component shifts throughout a pulsational phase, the static emission causes the apparent reversal in the absorption dips. 

\begin{figure*}
\caption{\label{cool_lines} Time-variable line profiles of our FG-type stars. The shown lines are the H$\alpha$ (solid thin red), the low-excitation \ion{Ba}{II} $\lambda614.17$ (solid thick blue), the high-excitation \ion{Si}{II} $\lambda634.71$ (dotted black), and the \ion{Fe}{I} $\lambda639.36$ (dashed green). Note the variable H$\alpha$ due to infalling (outflowing for HD271182 in 2017) shocked gas, as well as the variable asymmetry
and strengthening of \ion{Ba}{II}. The double-peaked \ion{Fe}{I} line of HD269953 arises from a disk that is shown to be detached from the hypergiant (see text).}\
\includegraphics[width=13cm,clip]{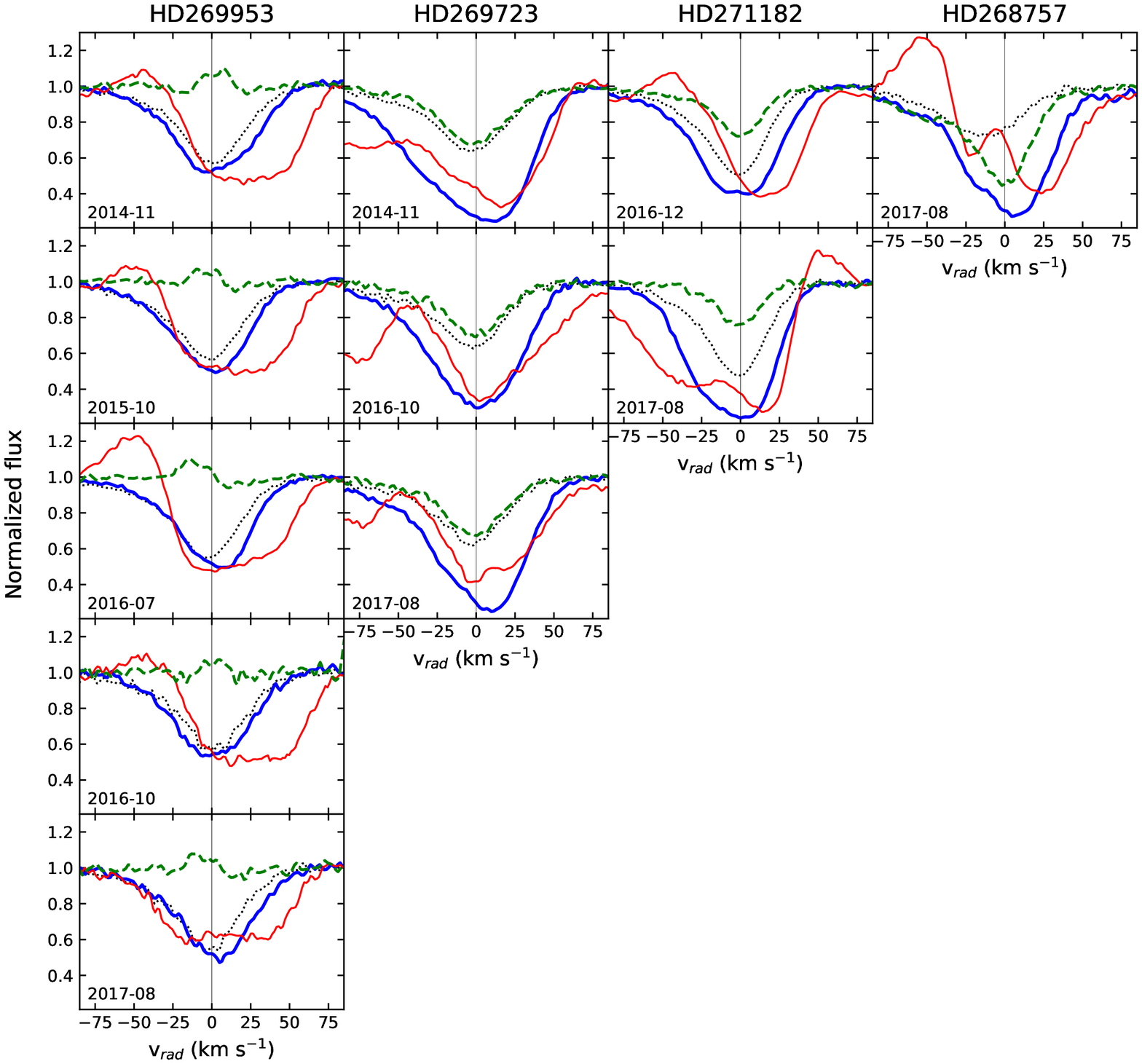}
\end{figure*}

In Fig. \ref{cool_lines}, we display the \ion{Ba}{II} $\lambda614.17$ line ($\chi_{\text{low}}=0.7$ eV) of the four late-type hypergiants, the asymmetric profile of which suggests the presence of a second, shortward-displaced component. The line was deblended with the IRAF routine \textit{splot} using two Gaussian profiles and their centroid velocities are given in Table \ref{late_vel}. Alongside we provide the velocities of the emission-filling H$\alpha$ components that trace the ejection or infall of cooling gas (Fig. \ref{cool_lines}). For comparison with the frame of the photosphere, we also display in Fig. \ref{cool_lines} the lines of \ion{Fe}{I} $\lambda639.36$ ($\chi_{\text{low}}=2.4$ eV) and \ion{Si}{II} $\lambda634.71$ ($\chi_{\text{low}}=8.1$ eV). The \ion{Fe}{I} emission of HD269953 is separately discussed in Section \ref{hd269953}. Lines with a high excitation potential such as the \ion{Si}{II}, are generated in the deeper layers of the hypergiant and are less prone to mechanisms that can interfere with a symmetric profile or introduce emission \citep[e.g.][]{kloch16,kloch19}. The \ion{Si}{II} line of HD268757 displays a blueshift of $13$ km~s$^{-1}$ with respect to the systemic velocity. We failed to observe similar shifts in other photospheric lines of the star. It is likely that the particular line follows the outward motion of deep layer as part of a pulsational cycle \citep[e.g.][]{Kraus19}. The weak shifts in the \ion{Si}{II} line of HD269953 are within the uncertainty of the photospheric velocity frame (see Section \ref{hd269953}), whereas no measurable offsets are observed in the FEROS data of HD269723 and HD271182.

\begin{table}
    \caption{\label{late_vel} Velocities measured for the identified components of H$\alpha$ (emission) and \ion{Ba}{II} (absorption). The uncertainty in the measurements is 0.5 km~s$^{-1}$.}
    \centering
    \begin{tabular}{lccc}
    \hline
    \multirow{2}{*}{Star} & \multirow{2}{*}{Date} & H$\alpha$ em. & \ion{Ba}{II} abs. \\
    & & (km~s$^{-1}$) & (km~s$^{-1}$) \\
    \hline 
    HD268757 & 2017-08 & $-$55, $-$5 & $-19$, $+$13 \\
    HD271182 & 2016-12 & $-$45 & $-$15, $+$13\\
             & 2017-08 & $-$10, $+$50 & $-$20, $+$8 \\
    HD269723 & 2014-11 & $-$45(?) & $-$22, $+$17 \\
             & 2016-10 & $-$42(?) & $-$  \\
             & 2017-08 & $-$47(?) & $-$22, $+$14 \\
    HD269953 & 2014-11 & $-$45 & $-$6, $+$24 \\ 
             & 2015-10 & $-$51 & $-$ \\
             & 2016-07 & $-$51 & $-$18, $+$10 \\
             & 2016-10 & $-$44 & $-$8, $+$14 \\
             & 2017-08 &  $-$ & $-$17, $+$7 \\
    \hline
    \end{tabular}
\end{table}

Although a thorough insight into the dynamics of the mass loss requires a continuous high-resolution monitoring, an interplay between the stellar feedback and the circumstellar reservoir can be preliminarily assessed. For HD271182, the inverse P Cygni of H$\alpha$ in 2016 captures the collapse of the upper atmosphere, similar as in the pre-outburst phases of $\rho$ Cas \citep{lob03,Kraus19}. Within a year, when the star is found to be in a hotter state (Table \ref{final_param}), outflowing gas accompanied by an increase in the density of the \ion{Ba}{II}-forming outer layer is observed. We therefore speculate that a launch of ejecta was powered by an energetic event that took place in early/mid 2017. Supportive of an enhanced mass-loss activity is also the light curve of the star (Fig. \ref{lc}), which shows fluctuations in the magnitude that exceed the baseline amplitude and indicates changes in the atmospheric opacity. The three-year data of HD269723 demonstrate a possible, almost static, violet-displaced H$\alpha$ component that grows within a broad absorption core of varying strength. Amongst the sample stars, HD269723 displays the largest separation between the \ion{Ba}{II} components. The asymmetry in the line diminishes in 2016, which can be explained by the attenuation of the circumstellar emission that intrudes the line. When the asymmetric profile is restored, in 2017, the H$\alpha$ core is found to be weakened due to underlying emission from recombining gas spanning the extended circumstellar volume. The spectroscopy of HD269953 tells a similar story; infalling gas gives rise to blueshifted H$\alpha$ emission that is intensified in July 2016. The component subsides entirely only in 2017, alongside the weakening of the broad core due to the envelope cooling. The observed \ion{Ba}{II} profile of HD269953 highlights the existence of a circumstellar layer, the velocity of which varies with time. In particular, the deep line core appears to be in anti-phase with the double-peaked \ion{Fe}{I} emission (Fig. \ref{cool_lines}) from the companion disk (see Section \ref{hd269953}), suggesting a coupling between the surroundings of the hypergiant and the orbital dynamics of the newly-identified system. Lastly, the single-epoch data of HD268757 expose two filling-in H$\alpha$ components, which trace collapsing gas at different regions of the upper atmosphere. Notably, a superimposed emission causing the observed splitting of \ion{Ba}{II} has a similar velocity with one of the H$\alpha$ components, implying line formation in a common layer within the envelope.

\section{The disk-hosting companion of HD269953}
\label{hd269953}

\begin{figure}
\caption{\label{HD269953_disk} Emission lines identified in the FEROS spectra of HD269953.}
\includegraphics[width=8.7cm,clip]{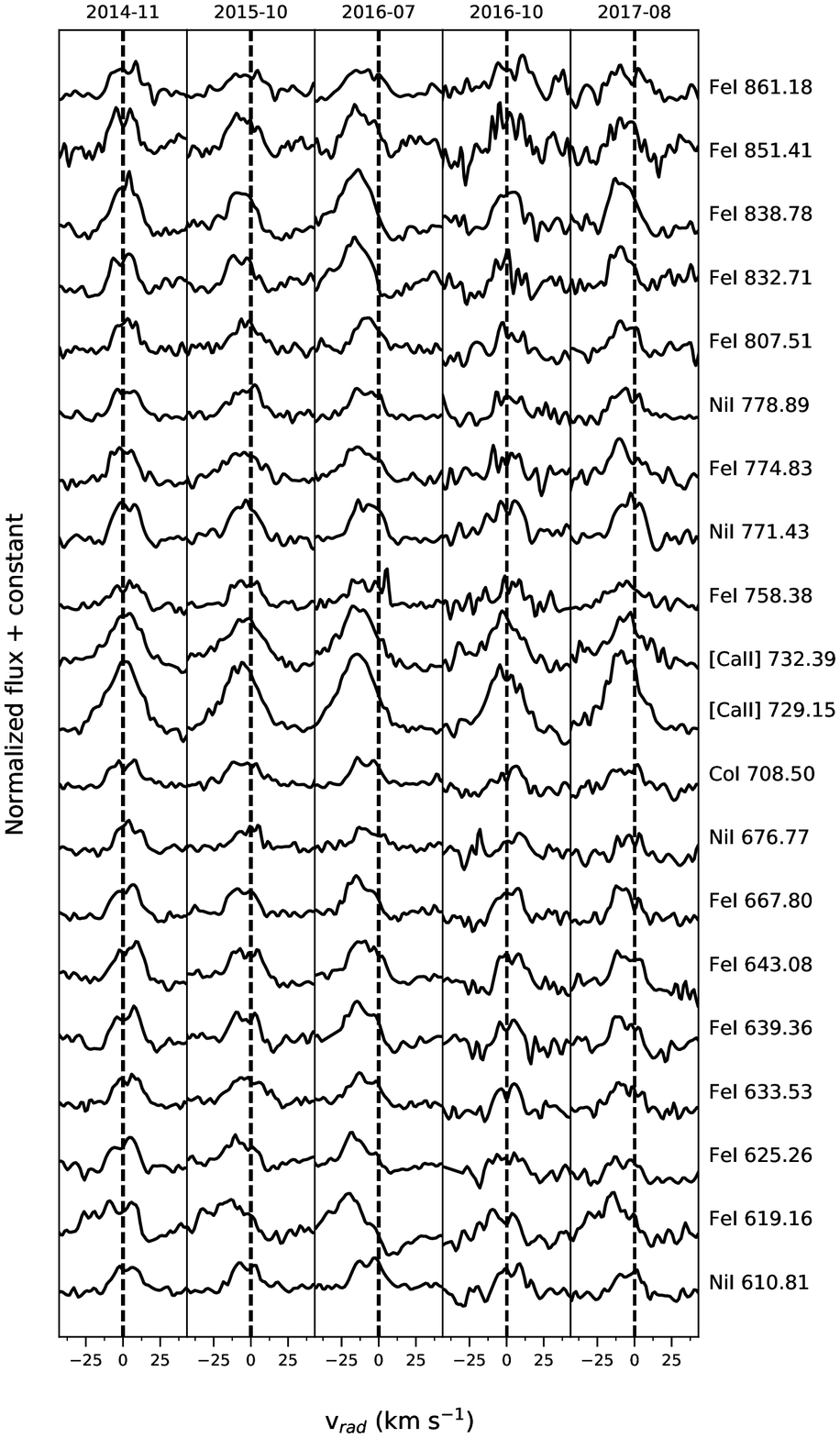}
\end{figure}

The hypergiant HD269953 stands out from the sample stars due to several lines found in emission. In Fig. \ref{HD269953_disk}, we collectively show these lines that were identified against the NIST database\footnote{https://physics.nist.gov/PhysRefData/ASD/lines\_form.html}. Neutral iron lines constitute the majority, being followed by lines of other metals such as \ion{Ni}{I} and \ion{Co}{I}. For several of the lines, a double-peaked profile can be resolved suggesting a component with disk-like geometry. The bluer of these lines capture changes in the ratio of red to blue component due to rotating inhomogeneities that are particularly evident when comparing the spectrum of 2014 and that of July 2016. Emission of [\ion{Ca}{II}] $\lambda\lambda729.2,732.4$ displays broad wings and lacks signs of resolved separation. The peak of the emission is within 5 km~s$^{-1}$ consistent with the centroid of the double-peaked lines, particularly tracing the strongest of the two components.

\begin{figure}
\caption{\label{HD269953_photosph} Mean velocity measurements from the systematic fit of neutral (blue dots) and single-ionized (red squares) absorption lines of \ion{Fe}, \ion{Ti}, \ion{Cr}, and \ion{Ba}, in the five spectra of HD269953. The light and dark shaded regions indicate the $\pm1\upsigma$ uncertainty around the centroid velocities of the double-peaked (disk) and [\ion{Ca}{II}] emission, respectively.}
\includegraphics[width=8.7cm,clip]{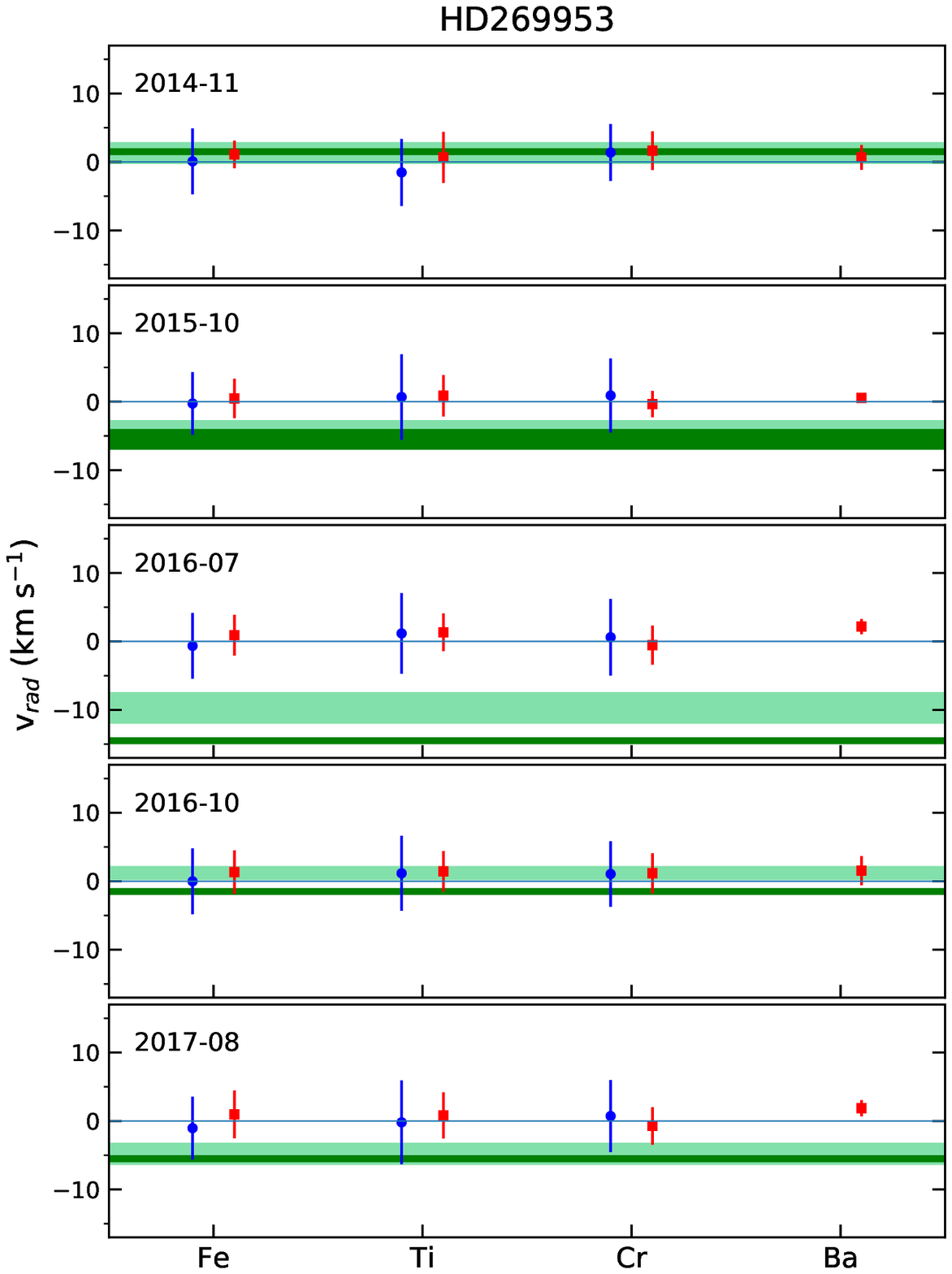}
\end{figure}

To explore the origin of the emission within the environment of HD269953, we examined the velocity of these lines within the frame of photospheric absorption. On average per spectrum, we identified against VALD and measured 180 photospheric lines of \ion{Fe}{}, 60 lines of \ion{Ti}{}, 60 lines of \ion{Cr}{}, and five lines of \ion{Ba}{II}. The mean values and uncertainties are shown in Fig. \ref{HD269953_photosph}. The lines of \ion{Ba}{II} show displacements due to their circumstellar origin.  On the same plot, we display the centroid velocities within $\pm1\upsigma$ uncertainty for both the double-peaked and [\ion{Ca}{II}] emission. The spectra of 2015, 2017 and, greatly that of July 2016, indicate deviation between the proposed disk component and the absorption lines of the hypergiant. The kinematics of the disk appears to be detached from the line frame of the photosphere, thus suggesting a component with orbiting motion.

\cite{2013A&A...558A..17O} undertook $K-$band spectroscopy of HD269953 and reported emission of the first overtone band heads for the CO molecule. The authors suggested that the emission arises in a ring that is part of a circumstellar disk, the composition of which reflects that of the stellar surface. Moreover, by measuring a ratio $^{12}$C/$^{13}$C $=10\pm1$, it was concluded that the star is in a post-RSG phase exposing processed material that has been dredged up from the core \citep{2009A&A...494..253K}. Using the stellar parameters from \cite{deJag98}, the star was illustrated to lie coolward of the evolutionary track of M$_{\text{ini}}=30$ M$_\odot$. In the present study, we update the parameters of HD269953 corresponding to a hotter and more luminous star than previously assumed. We currently provide an excellent agreement between the value of $^{12}$C/$^{13}$C measured by \cite{2013A&A...558A..17O} and the predicted value for a star with M$_{\text{ini}}=40$ M$_\odot$. 

Our recently obtained $K-$band spectroscopy of HD269953 reveals that the velocity of the photospheric lines is not in agreement with that of the CO band heads, implying that the molecular emission does not originate in the circumstellar environment of the hypergiant (Kraus et al. in prep.). A possible scenario is that it rises from the above identified companion disk. Essentially then, the double-peaked optical lines trace an accretion disk, which consists of processed material from the surface of HD269953 delivered via a past or ongoing Roche-lobe overflow. Alternatively, a circumbinary disk would be the source of CO further justifying the large line-of-sight emitting area calculated by \cite{2013A&A...558A..17O}.

\section{Discussion}
\label{discuss}

\begin{figure*}
\centering
\caption{\label{hr_diag} Hertzsprung–Russell diagram of massive star evolution at advanced phases. We show the refined position of the eight studied stars (red circles). The parameters and uncertainties were inferred as the average of different features of a single-epoch FEROS spectrum (small markersize) or out of the multi-epoch parameters (large markersize). The Humphreys-Davidson limit is displayed with thick solid line. Overplotted are the known YHGs (triangles) and LBVs (squares) from the literature \citep[][and references therein]{deJag98,2017AJ....153..125S}. The quiescence and eruptive states of a star are joined by a dotted horizontal line. Evolutionary models at $Z=0.014$ (solid) and $Z=0.002$ (dashed) assuming a rotation rate $\upsilon_{\text{ini}}/\upsilon_{\text{crit}}=0.4$ are taken from \protect\cite{2012A&A...537A.146E} and \protect\cite{2013A&A...558A.103G}, respectively.}
\includegraphics[width=14.5cm,clip]{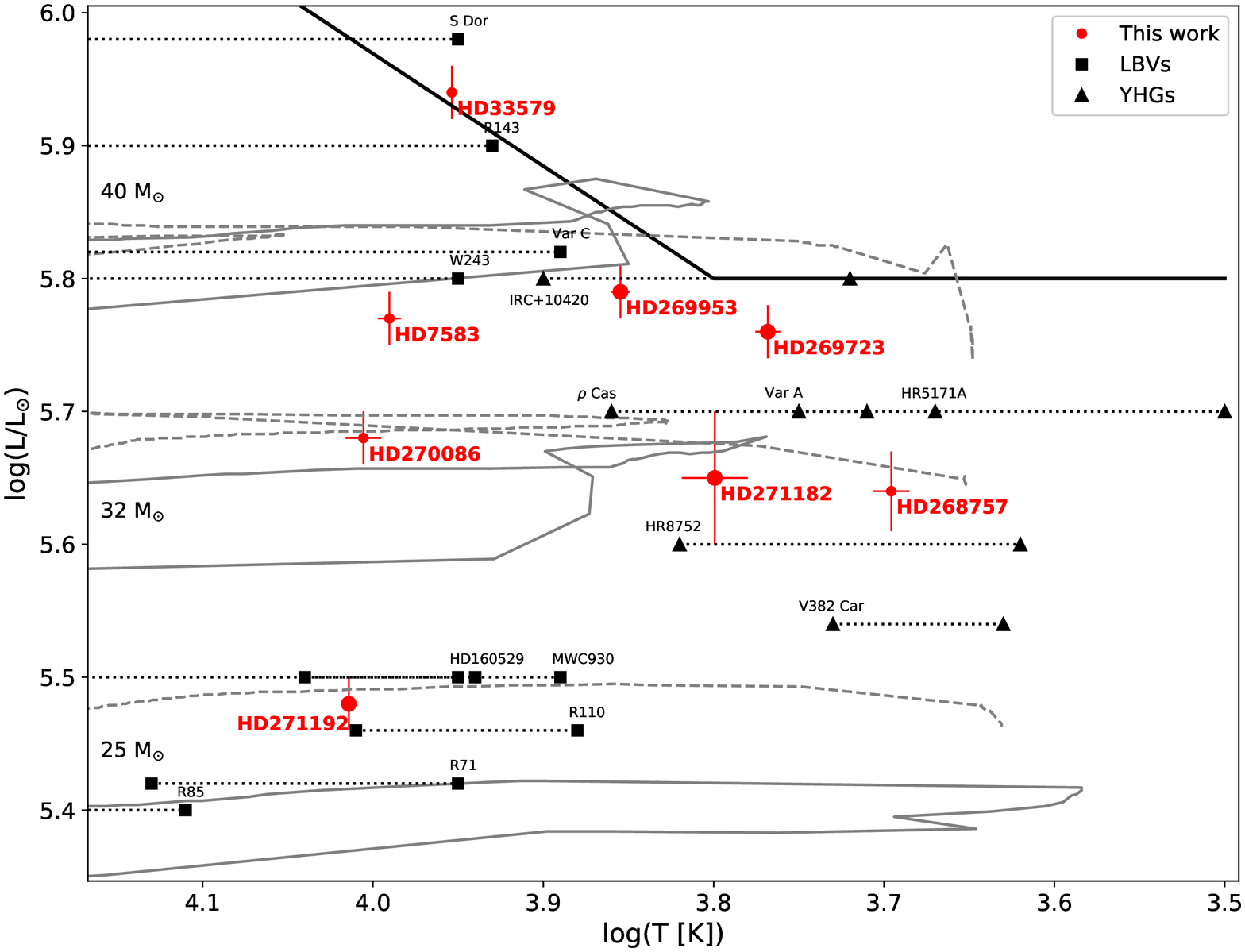}
\end{figure*}

The stellar parameters extracted from the echelle spectra and the multi-band photometry confine the location of the eight stars on the evolutionary diagram in Fig. \ref{hr_diag}. We display the single-epoch parameters (small markersize) and the mean values of the multi-epoch data (large markersize). The standard error of the mean is adopted as the uncertainty of the latter points, with the errorbars indicating the 2$\sigma$ values. Overplotted are the currently available sets of evolutionary models from the Geneva group at $Z=0.014$ \citep{2012A&A...537A.146E} and $Z=0.002$ \citep{2013A&A...558A.103G}, both accounting for initial rotation at the 40 per cent of the critical velocity. For comparison, we display the location of known eruptive YHGs and LBVs, when captured in both quiescent state and outburst \citep{deJag98,2017AJ....153..125S}. 

The initial masses of the A-type stars range from 27$-$30 M$_{\odot}$ to higher than 40 M$_{\odot}$. Collectively, these stars display characteristics typical of early-A supergiants, which include the P Cygni profile in H$\alpha$ and the weak infrared excess due to winds. Evident also is the absence of dust in their SEDs that would manifest as the relic of a prior cool or extreme mass-loss phase. As three of these stars occupy the same location on the evolutionary diagram as LBVs in their cool state, it is tempting to suggest our sample as such. Nevertheless, the FEROS spectra lack emission of critical forbidden lines, at the same time that the time series fail to display evidence of an eruptive behavior. A classification as low-luminosity LBVs is therefore discouraged. On the other hand, not all post-RSGs that have crossed the YV do necessarily appear as LBVs. For example, the class of $\alpha$ Cyg variables shows non-radial pulsations, which are excited only in stellar models evolving bluewards from a previous RSG phase \citep{2013MNRAS.433.1246S}. With the exception of HD33579, the ASAS$-$3 data of the A-type sample lack any measurable periodic modulation. To properly establish or argue against a classification as LBVs or $\alpha$ Cyg variables, however, a multi-epoch spectroscopy is necessary \citep[e.g.][]{2015A&A...581A..75K,2018A&A...613A..33C}. \cite{1991A&A...245..593H} suggested that both HD33579 and HD7583 are in a pre-RSG state by comparing to other A-type supergiants in the LMC and SMC that were shown to display abnormally strong hydrogen lines in their spectra. The latter finding was considered to be a non-LTE effect of the enhanced helium in the atmospheres, which differentiates post-RSGs from redward evolving stars. A same conclusion on the evolutionary state of HD33579 was given by \cite{deJag98} based on the solar-type abundances and high mass of the star. Throughout its recorded photometric history, HD33579 has shown no evidence of eruptions typical of YHGs, which led to its characterization as a remarkably stable star \citep{2000A&A...353..163N}.

We conclude that, the collective information can not support a classification of the four A-type stars other than as luminous and coolward evolving supergiants that possess steady mass-loss outflows.\\

The high luminosity of HD269953 is marginally of a star that does not experience a RSG phase and therefore, drawing conclusion on the evolutionary state of the star is a delicate task. In search of short-term variability, \cite{2019ApJ...878..155D} analyzed the high-cadence time series of the star obtained by the Transiting Exoplanet Survey Satellite (TESS). The range 1.1$-$1.6 d$^{-1}$ containing the strongest identified peaks was found to be higher by an order of magnitude compared to that of the other studied yellow supergiants. This fact led to the conclusion that HD269953 exhibits variability that is not typical for a pre-RSG star, suggesting so its post-RSG nature. At the same time, \cite{2019A&A...631A..48V} noted that the amplitudes and quasi-periods of pulsations in YHGs are found to increase during the redward loop of the stars and vice versa. Assuming that HD269953 is a post-RSG, the low photometric temperature of 4\,900 K adopted by \cite{2019ApJ...878..155D} would then raise expectations for a prominent variability in the photometry that, however, is not supported from the ASAS$-$3 data. On the other hand, the current spectroscopically measured $T_{\text{eff}}$ can well support the conclusions of \cite{2019A&A...631A..48V}. With a temperature exceeding $7\,000$ K, the star may have already crossed the cool border of YV \citep[based on the definition of][]{2012A&A...546A.105N}, potentially explaining the absence of strong peak frequencies in the power spectrum when compared to our rest FG sample (Fig. \ref{fourier}). In line with this scenario, the optical photometry of IRC+10420, a post-RSG that lies within the YV, is also shown to be invariable within an accuracy of 0.1 mag \citep{1996MNRAS.280.1062O}. The gaseous circumstellar environment of the latter verifies that the star survived a phase of severe mass-loss episodes \citep{2010AJ....140..339T}. Accordingly, the previously reported CO emission of HD269953 may trace the ejecta that has built up from a preceding phase as a YHG. The role of the identified companion in promoting or sustaining the phases of enhanced mass loss of the star remains to be resolved.

The baseline photometry of HD268757, HD271182, and HD269723, is outlined by cycles with uneven peaks and dims with variable depths. The strong pulsational activity mirrors the temperature oscillations of few hundreds of K throughout the different epochs, being most significant for HD271182 (Table \ref{final_param}). The longest cycles are identified for HD269723, especially prior JD2453600 when the ASAS$-$3 data suggest a possible period exceeding 1\,500 d (Fig. \ref{lc}). Post this date, the period is shortened to approximately 1\,200 d. Evident is also the decrease in the amplitude of variability with time. In the frame of the variable YHGs, the decrease in the pulsational parameters of the stars is linked to a blue-loop evolution \citep{2019A&A...631A..48V}, though a similar conclusion for HD269723 requires the inspection of colour time series. The three FEROS spectra of HD269723 spanning 2014$-$2017 do not show significant changes in the temperature, however, they capture an enhancement, attenuation, and partial recovery to the state of 2014, of the circumstellar \ion{Ba}{II} (Fig. \ref{cool_lines}). The timescale of this pattern is comparable to the pulsational period, suggesting that the energy that is deposited into the atmosphere by the pulsating photosphere modulates the opacity of the envelope. The cool circumstellar dust (Fig. \ref{cool_seds}) is evidence of the effective outflow of gas, although, it can also indicate the recent departure of HD269723 from the RSG phase, given that the star evolves faster than the less massive stars of the sample showing weaker or no infrared excess. 

The ASAS$-$3 brings to light the case of a bona fide YHG in the LMC, which undergoes an episodic event of enhanced mass loss. The outburst of HD271182, which manifests as a dim in the lightcurve with an amplitude 0.4 mag, follows the increase in the $V-$band flux to the brightest point of the nine-year data (Fig. \ref{lc}). The duration of the dim, regarded here as the time interval from the point of prior brightening to that of recovery to baseline level, is 300 d.  When comparing to both the Millenium outburst \citep{lob03} and the eruptive event of 2013 \citep{Kraus19} of $\rho$ Cas, the photometric data indicate that the LMC hypergiant underwent in late 2008 a less energetic eruption. Both amplitudes of the dim and of the pulsations are smaller than the respective ones of $\rho$ Cas. Nevertheless in both stars, the dims are followed by the characteristic rise in the brightness. The increase in the $V-$band brightness of $\rho$ Cas prior the Millenium outburst was followed by the collapse of the deeper photosphere, as this was confirmed by the strong redshift of the \ion{Fe}{I} lines \citep{lob03}. The rise in the local temperature to the range $7\,000-8\,000$ K, where gas is shown to be highly expandable, inserted a significant amount of energy that was released during the recombination cooling and upon the synchronous pulsational decompression. Although we lack spectroscopy that pairs the photometry of HD271182 during the episode, the significant brightening within an evident frame of pulsations can favor the above scenario. Because the envelopes of stars in the post-RSG phase are highly unstable, the deep photometric event might have caused an ejection of mass, which is still in orbit around the star. Indication for circumstellar gas is provided by the detected [\ion{Ca}{II}] emission lines (Fig. \ref{overview}). On the other hand, the amount of mass that had been ejected during that event seems to be insufficient for the formation of detectable amounts of dust as in HD269953 (Fig. \ref{cool_seds}). We suggest that HD271182 is an extragalactic post-RSG and a ``modest'' analogue to $\rho$ Cas.

The nature of HD268757 is questionable and any conclusion should be currently considered with caution. The time series of the star prior JD2453100 is similar to the quiescent phase of HD271182, namely a multi-period pulsation pattern with cycles lasting hundreds of days and a mean amplitude 0.17 mag. However, the sudden drops in the magnitude, with no sign of an imminent event, do not resemble that of HD271182 and neither those of $\rho$ Cas, the latter two being characterized by a steep temperature and magnitude increase preceding a sharp decline in $V-$band flux. We speculate that the coolest star of our sample is member of an eccentric eclipsing binary system with a period exceeding the nine-year window of ASAS$-$3.
The investigation of the spectrum of HD268757 identifies broad features, which are centered shortward of the rest frame by 30$-$40 km~s$^{-1}$, and that could be ascribed to a companion (see Fig. \ref{cool_lines}). Nevertheless, our attempt to superimpose model spectra of stellar components with various temperatures failed to give a satisfactory fit for several lines. On the other hand, contribution from different sources of radiation e.g. a disk surrounding the companion, would essentially fill several of the features of the unidentified star, similarly to what is seen for HD269953. The partial, post-eclipse recovery in the brightness of HD268757 in 2009 (see Fig. \ref{lc}), could be explained by a dust component in the environment of the hypergiant that prolongs the obscuration of the companion. The dusty surroundings of HD268757 are confirmed by the excess in the mid-infrared (Fig. \ref{cool_seds}). Alternatively, the post-eclipse decay in the $V-$band magnitude may denote the formation of an opaque shell due to dynamical conditions being met at periastron. We conclude that a proper classification of HD268757 requires more spectrophotometric data, although the long period of the system prevents an upcoming discussion on the fundamental parameters of the components.

\section{Conclusions}
\label{summary}

The identification of post-RSGs is a delicate process that requires the systematic review of various observational signs. Hence several counterparts, especially from the puzzling class of YHGs, are kept in the shadow and others, can be subject of misclassification. In the current study, we focused on eight evolved hypergiants in the LMC and SMC with the goal of refining their properties and update their evolutionary status. Our methodology combines the updated stellar parameters, spectroscopic diagnostics of atmospheric instability and enhanced mass-loss activity, and the long-term photometric monitoring in the $V$ band from ASAS$-$3. 

The temperatures of the stars were measured from our analysis of high-resolution spectroscopy with FEROS. For this purpose, we built a calibration scheme that uses depth ratios of selected metal lines and is insensitive to the extreme broadening effects that characterize the studied stellar types. The spectroscopic temperatures were used to fit the SEDs and constrain the stellar luminosities and extinction. By conducting Fourier analysis of the time series, we report peak frequencies for five stars showing multi-periodic signals.

The four A-type stars are shown to span a narrow range of temperatures $9\,000-10\,500$ K and display a similar spectrophotometric picture, which includes the P Cygni in H$\alpha$, the weak infrared excess due to winds, and the lack of circumstellar dust that could stand as remnant from a prior phase of enhanced mass loss. The stars also fail to show signatures typically seen in low-luminosity LBVs that are in a blueward evolution. With three of these objects lacking measurable periodic signals, a classification as $\alpha$ Cyg variables is also discouraged, though not excluded. With characteristics typical of early-A supergiants, the four stars are suggested here to be post-main sequence stars evolving redwards. Exceptionally, the bright HD33579 is situated on the HD limit that naturally halts its transition to the late spectral types. The high luminosity to mass ratio of the star predetermines a high probability for instabilities occurring in its outer envelope, and the multiple identified periods, spreading from a few tens to a few hundreds of days, could manifest as precursors for upcoming eruptive events.

The strong macroturbulent broadening characterizes the FEROS spectra of the four FG-type stars, being ascribed  to the large-scale atmospheric motion. The stars, which are assigned temperatures  $5\,000-7\,000$ K and luminosities exceeding $\log(L/L_{\odot})\sim 5.6$, manifest themselves through the energetic processes that take place in their upper atmosphere. Their H$\alpha$ is filled by recombining gas that follows infalling or outflow motion. Synchronously, a split low-excitation \ion{Ba}{II} line traces the deposition of circumstellar gas and is modulated by the dynamics of H$\alpha$. A multi-period pulsation pattern is identified for HD268757, HD271182, and HD269723. For the latter two stars, the activity mirrors the fluctuations in the spectroscopic temperature assessed from the FEROS data at different epochs. The longest pulsation cycles, exceeding 1\,000 d in period, characterize the dusty hypergiant HD269723. The three-epoch spectroscopy of the star captures phases of attenuation and replenishment of the circumstellar envelope within a time frame comparable to the pulsational period. Together with the sign of surrounding cool dust, the star is suggested to have recently departed from a RSG phase and exhibits dynamical instability.

The high-resolution spectroscopy of HD269953 uncovers a double-peaked emission profile for several lines of neutral metals. The emission component is shown to be decoupled from the absorption line frame of the hypergiant, suggesting an orbiting disk-hosting companion. A disk of processed material is in line with the results of previous observations on the presence and chemistry of CO molecular gas in the environment of HD269953. The follow-up monitoring of the target is essential for assembling the radial velocity curves. The accurate determination of the dynamical mass of a YHG will set a vital constraint on the physics of this transient phase. Star HD268757 is also proposed to be embedded within a binary, based on the time-series photometry of the star that is characteristic of an eccentric and long-term eclipsing system. The coolest YHG of our sample further shows evidence of separate infalling ejecta complemented by a split \ion{Ba}{II} line and cool circumstellar dust.

The pulsationally variable timeline of HD271182 is corrupted by an energetic event, which manifests as a photometric dim of amplitude 0.4 mag. Alongside, the two FEROS spectra capture phases prior and post an episodic ejection of gas that took place in early/mid 2017. Paired with the location of the star on the evolutionary diagram, the data suggest that HD271182 is a ``modest'' analogue to $\rho$ Cas that undergoes less dramatic episodes compared to the Galactic YHG. The luminosity of the star, $\log(L/L_{\odot})= 5.6$, sets a threshold for stars exhibiting a post-RSG evolution in the LMC, based on  the currently known sample of YHGs in the Clouds. To date, this value exceeds luminosities of YHGs in the Galaxy (e.g. of HR8752 and V382 Car), possibly linking to the auxiliary role of metallicity in unbinding the envelope of less massive RSGs but also, in the severity of the YHG eruptions thereafter.

\section*{Acknowledgements}

We thank the anonymous referee for the valuable comments that helped us to improve the quality of the paper. M. Kourniotis and M. Kraus acknowledge financial support from the Czech Science foundation GA\v{C}R under grant numbers 19-15008S and 20-00150S, respectively. GM acknowledges funding support from the European Research Council (ERC) under the European Union’s Horizon 2020 research and innovation programme (Grant agreement No. 772086). The Astronomical Institute of the Czech Academy of Sciences is supported by the project RVO:67985815. This project has also received funding from the European Union's Framework Programme for Research and Innovation Horizon 2020 (2014-2020) under the Marie Sk\l{}odowska-Curie Grant Agreement No. 823734. M. Kourniotis thanks Alceste Bonanos for providing comments on the manuscript. The study is based on observations collected with the MPG 2.2m telescope at the European Southern Observatory (La Silla, Chile) under programmes 094.A-9029(D), 096.A-9039(A), 097.A-9039(C),  098.A-9039(C), and 099.A-9039(C). The observations obtained with the MPG 2.2m telescope were supported by
the Ministry of Education, Youth and Sports project - LG14013 (Tycho
Brahe: Supporting Ground-based Astronomical Observations). We would like
to thank the observers (S. Ehlerov\'a,  A. Kawka, P. Kab\'ath, S. Vennes, \& L. Zychov\'a) for obtaining the data. This research has made use of the SIMBAD database, operated at CDS, Strasbourg, France. 

\section*{Data availability}

The data underlying this article will be shared on reasonable request to the corresponding author.

\bibliographystyle{mnras}
\bibliography{kourniotis2021}

\bsp
\label{lastpage}
\end{document}